\documentclass[twocolumn]{aastex631}
\usepackage{hyperref}

\usepackage[hang,flushmargin]{footmisc} 
\usepackage[T4, T1]{fontenc}
\usepackage{amsmath}	
\usepackage{amssymb}	
\usepackage{booktabs} 

\newcommand{\vida}[1]{\textcolor{black}{#1}}

\shorttitle{Dual AGNs in {\sc Romulus25} Simulation}
\shortauthors{Saeedzadeh, Babul,Mukherjee, Tremmel, Quinn, Mayer}

\begin{document}

\title{ Dual AGNs: Precursors of Binary Supermassive Black Hole Formation and Mergers}

   \author[0009-0000-7559-7962]{Vida Saeedzadeh}
   \email{vidasaeedzadeh@uvic.ca}
    \affiliation{Department of Physics and Astronomy, University of Victoria, 3800 Finnerty Road, Victoria, BC, V8P 1A1, Canada}

    \author[0000-0003-1746-9529]{Arif Babul}
    \affiliation{Department of Physics and Astronomy, University of Victoria, 3800 Finnerty Road, Victoria, BC, V8P 1A1, Canada}
    \affiliation{Infosys Visiting Chair Professor, Department of Physics, Indian Institute of Science, Bangalore 560012, India}
    \affiliation{Leverhulme Visiting Professor, Institute for Astronomy, University of Edinburgh, Royal Observatory, Blackford Hill, Edinburgh EH9 3HI, United Kingdom}
    \author[0000-0002-3373-5236]{ Suvodip Mukherjee}
    \affiliation{Department of Astronomy \& Astrophysics, Tata Institute of Fundamental Research, 1, Homi Bhabha Road, Colaba, Mumbai 400005, India}
    \author[0000-0002-4353-0306]{Michael Tremmel} 
    \affiliation{School of Physics, University College Cork, College Road, Cork T12 K8AF, Ireland}
    \author[0000-0001-5510-2803]{Thomas R. Quinn}
    \affiliation{Astronomy Department, University of Washington, Box 351580, Seattle, WA, 98195-1580, USA}
    \author[0000-0002-7078-2074]{Lucio Mayer}
    \affiliation{Institute for Computational Science, University of Zürich, Winterthurerstrasse 190, 8057 Zürich, Switzerland}    



\begin{abstract}

The presence of dual active galactic nuclei (AGN) on scales of a few tens of kpc can be used to study merger-induced accretion on supermassive black holes (SMBHs) and offer insights about SMBH mergers, using dual AGNs as merger precursors. This study uses the {\sc Romulus25} cosmological simulation to investigate the properties and evolution of dual AGNs. We first analyze the properties of AGNs ($L_{bol} > 10^{43} \rm $ erg s$^{-1}$) and their neighboring SMBHs (any SMBHs closer than 30 pkpc to an AGN) at $z \leq 2$. This is our underlying population. We then applied the luminosity threshold of $L_{bol} > 10^{43} $ erg s$^{-1}$ to the neighboring SMBHs thereby identifying dual and multiple AGNs. Our findings indicate an increase in the number of both single and dual AGNs from lower to higher redshifts. We also find that the number of dual AGNs with separations of 0.5-4 kpc is twice the number of duals with separations of 4-30 kpc. All dual AGNs in our sample resulted from major mergers. Compared to single AGNs, duals have a lower black hole-to-halo mass ratio. We found that the properties of dual AGN host halos, including halo mass, stellar mass, star formation rate (SFR), and gas mass, are generally consistent with those of single AGN halos, albeit tending towards the higher end of their respective property ranges. Our analysis uncovered a diverse array of evolutionary patterns among dual AGNs, including rapidly evolving systems, slower ones, and instances where SMBH mergers are ineffective.

\end{abstract}

\keywords{ active galaxies - active galactic nuclei - black hole physics - computational methods}


\section{Introduction} \label{sec:intro}

The currently favored hierarchical structure formation paradigm, wherein the cosmic structure is built up via successive mergers of systems of increasingly larger size, essentially means that SMBH-SMBH mergers will be ubiquitous over cosmic time, especially since most galaxies are thought to host SMBHs \citep[e.g.][]{kormendy2013review}. The path from galaxy-galaxy mergers to the eventual SMBH-SMBH mergers involves a sequence of processes spanning a large dynamic range of spatial scales. In some cases, two SMBHs will pass through a phase where both are accreting and radiating efficiently \citep[e.g.][]{diMatteo2005,hopkins2008A}, thereby appearing as close pairs of luminous AGNs \cite[e.g.][]{gerke2007,comerford2009}. Dual AGN events can be used to study accretion on SMBHs and as precursors of SMBH-SMBH mergers \citep[e.g.][]{saeedzadeh2023bbh}.  They have also been proposed to be indicators of galaxy mergers \citep{comerford2009}.

Over the past decade, significant attention has been devoted to the detection and study of dual AGNs. Observationally, many dual AGNs have been detected serendipitously. However, recent systematic searches are beginning to address the statistical properties of these systems \citep{deRosa2019review}, focusing on both their occurrence and properties. These detections primarily cover low redshifts, but recent efforts by several groups are extending these detections to high-redshift close pairs using novel observational techniques \citep{shen2019,shen2021,hwang2020,silverman2020,chen2022,mannucci2022}. A small number of multiple AGN systems have also been reported in the literature \citep[e.g.][]{djorgovski2007,deRosa2015,hennawi2015,liu2019,pfeifle2019}, with separations ranging from a few tens to hundreds of kpc.

Given the recent observations and the increasing detection of dual AGNs, there is a growing need for a sample of simulated counterparts to understand the observed sample and its astrophysical implications. Within the realm of idealized galaxy-merger simulations, several studies have focused on AGNs turning on at various pair separations. They also investigate the impact of galaxy merger parameters, including host galaxy mass ratio and morphology on turning on an AGN \citep[e.g.][]{vanWassenhove2012,blecha2013,capelo2017} and black hole merger timescales \citep[e.g.][]{mayer2017bbhmerger}. Idealized merger simulations can resolve sub-kpc scales, which are essential for tracking the dynamics of SMBHs. Nonetheless, these simulations cannot provide predictions about the fraction of AGN pairs in comparison to the total number of AGNs in the universe.

On the other hand, cosmological simulations, which have
lower resolution than idealized simulations, can in principle, identify
the incidence of dual AGN and shed light on their origins.
For example, \citet{steinborn2016} and \citet{chen2022} analyzed the differences between dual and offset AGN. \citet{volonteri2016} and \citet{Ricarte_2021} examined dual AGN in the context of wandering SMBHs, which are the population of SMBHs that do not settle in the galaxy center and, therefore, are not merging with the central SMBH. \citet{rosasGuevara2019} investigated the abundance of dual AGN as a function of redshift and confirmed that non-simultaneous accretion on SMBHs decreases the detection probability. \citet{volonteri2022dualagn}  investigated the connection between dual AGNs, galaxy mergers, and SMBH mergers. Their study was also expanded to include multiple AGNs.

Among the cosmological simulations mentioned above, very few have been able to produce sub-kpc dual AGNs. This is primarily because achieving sub-kpc spatial resolution, necessary to resolve such close pairs, is computationally expensive in a cosmological context. Additionally, in many simulations, SMBHs are pinned to the gravitational potential minimum to avoid artificial scattering of the SMBH \citep[e.g.][]{Crain.2009,Sijacki_2015MNRAS,dave2019simba}. This results in the rapid merging of two central SMBHs during a galaxy merger, precluding their capture at $ \sim \rm kpc$ separations.

{\sc Romulus} is a cosmological simulation that offers the advantages of large-scale simulations for studying the statistics of dual versus single AGN, while also providing high resolution capable of resolving sub-kpc scales. Additionally, the dynamical-friction modeling in {\sc Romulus} allows for one of the first studies of the evolution of AGN pairs and their activation in the context of cosmological simulations. In this paper, we are using {\sc Romulus25} which is a (25 $\rm cMpc)^3$ volume simulation with gas and dark matter mass resolution of $2.12\times 10^5 M_\odot$ and $3.39 \times 10^5 M_\odot$, respectively, and plummer equivalent spatial resolution of 0.25 kpc, which allows AGN pairs to be resolved at separations of few hundred parsecs, a few hundred Myrs after their host galaxies have merged.

In this paper, we investigate the properties of dual AGNs and their host galaxies in comparison with single AGNs. The paper is organized as follows: Section \ref{sec:romulus} provides a summary of the Romulus simulations and explains the detection and definition criteria used to select AGNs and dual/multiple AGNs. In Section \ref{sec:results}, we present the results of our analysis, starting from the analysis of AGNs and all their neighboring SMBHs to the properties of dual AGNs and their host halos in comparison with those of single AGNs, highlighting differences and patterns. We also explore the evolution of dual AGNs over time. Finally, in Section \ref{sec:conclusion}, we summarize our findings.

\section{Methods}

\subsection{Romulus Simulation}\label{sec:romulus}

In this study, we present results from analyzing the {\sc Romulus25} simulation. {\sc Romulus25} is a \vida{$\rm (25\ cMpc)^3$} cosmological volume simulation from the Romulus suite \citep{tremmel2017romulus,tremmel2019introducing,butsky2019ultraviolet,jung2022massive,saeedzadeh2023cgm}.   

The simulation was run using 
the Tree+Smoothed Particle Hydrodynamics (Tree+SPH) code CHaNGa \citep{menon2015adaptive,wadsley2017gasoline2}, with the Plummer equivalent gravitational force softening of 250 pc (or 350 pc spline kernel), a maximum SPH resolution of 70 pc, and gas and dark matter particle masses of $2.12\times 10^5 \rm\ M_\odot$ and $3.39 \times 10^5 \rm\ M_\odot$, respectively.  The background cosmology is a $\Lambda$CDM universe with cosmological parameters consistent with the Planck 2016 results \citep{Ade:2015xua}: $\Omega_{\rm m} = 0.309$, $\Omega_{\rm \Lambda} = 0.691$, $\Omega_{\rm b} = 0.0486$, $\rm H_{\rm 0} = 67.8\, {\rm km}\,{\rm s}^{-1} \rm Mpc^{-1}$, and $\sigma_{\rm 8} = 0.82$.

The full details about the {\sc Romulus25} simulation, including a thorough discussion of the hydrodynamics code and the specifics of the {\sc Romulus} galaxy formation model, the sub-grid physics incorporated therein, the various modeling choices made, and the simulation's many unique features, have been described in a number of published papers.  In the interest of brevity, we do not repeat this information here and instead refer interested readers to \citet{tremmel2015off,tremmel2017romulus,tremmel2019introducing,tremmel2020formation,sanchez2019not,butsky2019ultraviolet,chadayammuri2020fountains}; and \citet{jung2022massive}.  The latter especially offers a concise yet complete summary.  

There are, however, a few aspects of the {\sc Romulus25} simulation that are important to highlight as these are relevant to the present discussion. These pertain to the treatment of SMBH seeding, growth, and dynamical evolution in the {\sc Romulus25} \citep{tremmel2017romulus}.

\subsubsection{SMBH Seeding}

The {\sc Romulus25} simulation adopts a unique approach to SMBH seeding that diverges from the methodologies employed in many other cosmological simulations \citep[e.g.,][]{schaye2015eagle,weinberger2017tng,pillepich2018tng,dave2019simba}. In contrast to models that require a host halo or galaxy to exceed a specific mass threshold for the formation of a SMBH, in {\sc Romulus25} seeding of SMBHs depends only on the local gas properties \citep{tremmel2017romulus}. As a result, the SMBHs in {\sc Romulus25} can form in low mass halos and tend to form much earlier \citep[$z > $ 5, ][]{tremmel2017romulus}. Additionally, multiple SMBHs can arise in the same halo.

A gas particle converts into a SMBH seed if it meets the following criteria: (i) it is eligible and selected to form a star following a probabilistic process. (ii) it has very low metallicity ($ Z < 3 \times 10^{-4}$); (iii) its density is 15 times the threshold for star formation ( $ \geq 3\;m_p/{\rm cc}$), and (iv) its temperature is within the range of $9,500$ - $10,000\;$K.
This seeding mechanism resembles the direct collapse black hole scenario, where high temperatures and low metallicities inhibit fragmentation, enabling large gas clouds to directly collapse into an SMBH seed \citep{lodato2007mass,alexander2014rapid,natarajan2021new}.  In {\sc Romulus25}, the SMBHs are seeded with an initial mass of $\rm 10^6 \ M_\odot$ to ensure that they are always more massive than dark matter and star particles to mitigate spurious scattering events \citep{tremmel2015off}. 

The resulting SMBH occupation fraction at z=0 is consistent with current observations \citep{ricarte2019tracing}. Furthermore, the correlation between SMBH masses and their host galaxies' stellar masses follows the observed SMBH mass-stellar mass relation \citep{tremmel2017romulus,ricarte2019}.

\subsubsection{SMBH Dynamics and Mergers}\label{sec:romulus-bhmerger}

Another aspect where the {\sc Romulus25} SMBH model differs from other cosmological simulations lies in the treatment of SMBH dynamics. In contrast to several simulations that pin SMBHs at the centers of their host galaxies artificially \citep[e.g.][]{Crain.2009,Sijacki_2015MNRAS,dave2019simba}, {\sc Romulus25} accurately tracks the dynamical evolution of SMBHs down to sub-kpc scales, which is highly advantageous for this study. To accomplish this, the simulation implements a sub-grid model to represent the unresolved dynamical friction from stars and dark matter, which affects the SMBHs \citep{tremmel2015off}. This effect is calculated for each SMBH by assuming a local isotropic velocity distribution and applying Chandrasekhar's formula, integrating from the 90-degree deflection radius ($\rm r_{90}$) up to the gravitational softening length ($\epsilon_g$) of the SMBH. The resulting acceleration is

\begin{equation}
{\mathbf a}_{DF} = -4\pi G^2\; M_\bullet\; \rho(v < v_{BH})\; {\rm ln}\Lambda\;\frac{{\mathbf v}_{BH}}{v^3_{BH}},
\end{equation}

where $\rm ln\Lambda$ is defined as $\frac{\epsilon_g}{r_{90}}$, with ${\mathbf v}_{BH}$ denoting the SMBH's velocity relative to the local center of mass velocity of the closest 64 star and dark matter particles. Moreover, $\rho$ signifies the mass density, and G is the gravitational constant. 

For two SMBHs to merge, they need to be closer than two gravitational softening lengths (0.7 kpc) and have a low enough relative velocity that allows them to be gravitationally bound; that is, $\frac{1}{2} \; \Delta {\mathbf v}^2 < \Delta {\mathbf a} \cdot \Delta {\mathbf r}$, where $\Delta {\mathbf v}$ represents the velocity difference, $\Delta {\mathbf a}$ the acceleration difference between the black holes, and $\Delta {\mathbf r}$ the separation distance \citep{bellovary2011,tremmel2017romulus}\footnote{Note that there is a typographical error in the criterion for boundedness in \citet{tremmel2017romulus}.}. The criterion of two gravitational softening lengths is used because the simulation's precision in tracking the dynamics of the SMBH pair becomes less reliable below this threshold.

Upon merging, the resulting SMBH is assigned a velocity that conserves momentum, and its mass is the sum of its progenitors' masses. Mergers are one of the two processes through which SMBHs grow.
 
\subsubsection{SMBH Growth and Feedback}

Another mechanism for SMBH growth is through the accretion of gas. In the {\sc Romulus25} simulation, the rate of gas accretion is determined using a modified Bondi-Hoyle-Lyttleton formula (\citealt{bondi1952spherically},  for modifications see \citealt{tremmel2017romulus}) applied to the smoothed properties of the 32 closest gas particles:

\begin{equation}
    \dot{M}_\bullet = \alpha \times
    \begin{cases}
    \frac{\pi (G M_\bullet)^2 \rho_{\rm gas}}{(v_{bulk}^2 + c_s^2)^{3/2}} ~~~~~ \textrm{if} ~ v_{bulk} > v_{\theta} \\
    \\
    \frac{\pi (G M_\bullet)^2 \rho_{\rm gas} c_s}{(v_{\theta}^2 + c^2)^{2}} ~~~~ \textrm{if} ~ v_{bulk} < v_{\theta}
    \end{cases}, 
\end{equation}

where $\rho_{\rm gas}$ is the ambient gas density, $c_s$ is the ambient sound speed, $v_\theta$ is the local rotational velocity of surrounding gas, and $v_{bulk}$ is the bulk velocity relative to the SMBH.  These ambient quantities are calculated based on the 32 gas particles closest to the SMBH. The inclusion of $v_\theta$ and $v_{bulk}$ in this formulation is designed to address the original Bondi-Hoyle-Lyttleton approach's oversight of the gas's bulk movement and angular momentum. To correct for the suppression of the black hole accretion rate due to resolution effects, the coefficient $\alpha$ is employed, which is defined as

\begin{equation}
    \alpha = 
    \begin{cases}
    (\frac{n}{n_{th,*}})^2 ~~~ \textrm{if} ~~ n \geq n_{th,*}\\
    \\
    1 ~~~~~~~~~~~~ \textrm{if} ~~ n \leq n_{th,*}
    \end{cases},
\end{equation}

where $n_{th,*}$ is the star formation number density threshold ($0.2\; m_p/cc$).

The process of gas accretion onto a SMBH generates energy that is released into the black hole's surrounding area. In the {\sc Romulus25} simulation, this released energy is assumed to be electromagnetic, with a portion of it coupled to the nearby gas and contributing to its internal energy. The rate at which thermal energy is deposited is described by $\dot{E}_{\bullet,th} = \epsilon_r \epsilon_f \dot{M}_\bullet c^2,$ where $\epsilon_r$ represents the radiative efficiency (set at 10\%) and $\epsilon_f$ denotes the efficiency of coupling to the gas (set to 2\%). This thermal energy is isotropically distributed to the closest 32 gas particles based on the smoothing kernel. We refer readers to \citet{tremmel2017romulus} for further details.

\subsection{Halo catalogue and substructure definition}

In the Romulus simulations, halos are identified and processed using the Amiga Halo Finder \citep[AHF,][]{knebe2008relation,knollmann2009ahf}. Their evolution over time is tracked with TANGOS \citep{pontzen2018tangos}. 

Halos and subhalos form a hierarchical arrangement in which halos serve as the primary structures and the subhalos are nested within them. AHF employs an adaptive smoothing method to detect density peaks for identifying these structures. It identifies all gravitationally bound particles (including dark matter, gas, stars, and SMBHs) associated with each density peak and then moves to higher levels in the hierarchy to locate larger structures. After halos are identified, their centers are determined using the shrinking sphere method \citep{power2003inner}, which relies on the distribution of bound particles linked to each halo.

The masses of the halos ($M_\Delta$) are calculated by enclosing each halo center within a sphere of radius $R_\Delta$. This sphere is constructed so that the average density within it, $\left\langle{\rho_\mathrm{m,\Delta}}(z)\right\rangle$, matches $\Delta$ times the critical cosmological density, $\rho_\mathrm{crit}(z)$. Here, $\rho_\mathrm{crit}$ is critical density of the universe and $\Delta$ is a constant \citep[see, for example,][]{babul2002physical}. In this study, we refer to ($M_\mathrm{200}, R_\mathrm{200}$) which correspond to $\Delta = 200$.

The above prescription is only specifically for the halos.  For subhaloes, AHF tracks the local density profile relative to the distance from the peak's center. At some point, the external gravitational field becomes dominant, and correspondingly, the behavior of the density profile changes.   
The distance from the peak where this happens marks the size of the subhalo, and the mass enclosed is recorded as the subhalo's mass.

\subsection{Selection of dual/multiple AGNs}\label{sec:method-selection}

We perform our analysis at nine different outputs from $z$ = 0.05 to $z$ = 2  in increments of $\Delta z \sim$ 0.25. At each redshift, a detected AGN is defined as a SMBH with a bolometric luminosity  $\rm L_{bol} > 10^{43} $ erg s$^{-1}$. This threshold allows us to focus on AGNs that are sufficiently powerful to be identiﬁed observationally.

As mentioned in section \ref{sec:romulus} in the {\sc Romulus} black hole model, during the accretion process, thermal energy is injected isotropically into the surrounding gas particles, assuming a radiative eﬃciency of 10\% and a feedback coupling eﬃciency of 2\%. To determine the bolometric luminosities, we adopt the same radiative efficiency. Therefore, $L_{bol} = 0.1 \dot{M}_\bullet c^2$, where $\dot{M}_\bullet$ is mass accretion rate into the SMBH and $c$ is speed of light.

We begin by identifying SMBHs within a 30 pkpc from a detected AGN, labeling these as neighbor SMBHs. This 30 pkpc threshold is theoretically motivated by focusing on systems likely to interact or that are currently interacting. It also aligns with the typical separations used in observational searches for dual AGNs \citep[e.g.][]{comerford2013,silverman2020}. In this stage, the selection of neighbor SMBHs is not subject to specific mass or luminosity thresholds (note that our seed SMBH mass is $10^6 \, \rm M_\odot$; therefore all SMBHs necessarily have $M_\bullet > 10^6 \, \rm M_\odot$). These neighboring SMBHs may reside within the same halo as the AGN or in a separate halo. The combined set of AGNs and their neighbor SMBHs form the underlying population for our study.

To categorize AGNs as single, dual, or multiple, we investigate each detected AGN for neighboring SMBHs that are ``active''— defined as having $\rm L_{bol} > 10^{43} $ erg s$^{-1}$. An AGN is classified as ``single AGN'' if none of its neighboring SMBHs meet this luminosity criterion. Conversely, if one neighboring SMBH is active, we classify the AGN and its active neighbor as a ``dual AGN''. In scenarios where more than one neighboring SMBH is active, they are classified as ``multiple AGNs'', leading to the identification of triple and quadruple AGN configurations in our study. 

According to the procedure just described, two non-central SMBHs may be classified as a dual AGN, and systems with multiple AGNs may be counted as more than one dual. For example, in a scenario where three AGNs within a 30 kpc radius region meet the criteria, they would be counted as three separate dual AGNs. To prevent overcounting, we adopt a hierarchical approach, starting with the highest order of multiplets and working down to duals. We first identify clusters of four AGNs – the highest multiple for our reference luminosity and distance cuts – and remove them from the list. Then we proceed similarly with triplets, ultimately leaving us with ``pure'' dual AGNs.

In dual AGNs, the brighter AGN in the pair is referred to as the ``primary'' and the fainter one is called the ``secondary''. Primary and secondary AGNs are selected at the same output, i.e. at the same redshift. We identify the host galaxies of the SMBHs with Amiga Halo Finder (AHF). However, it should be noted that during the close encounters of galaxies, AHF is not always able to separate the merging systems well. This is especially the case for hosts that have undergone strong gas and stellar disruption. Finally, when tracing SMBH host galaxy properties back in time, we always follow the more massive progenitor if the SMBH of interest has gone through prior mergers.

\section{Results}\label{sec:results}

\begin{figure*}
    \centering
    \includegraphics[width= 1.\textwidth]{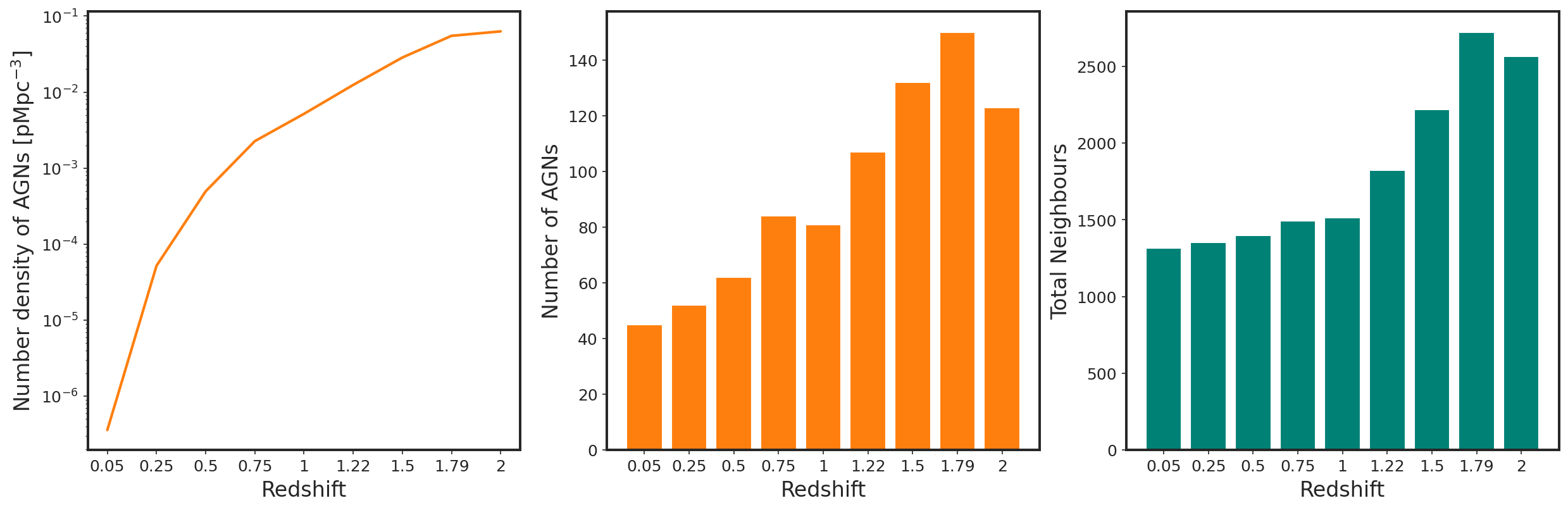}
    \caption{AGN number density, AGN counts and total number of neighboring SMBHs across redshifts. The first panel illustrates the number density of AGNs detected at redshifts under consideration. The middle and right panel presents the absolute number of AGNs and the total number of SMBH neighbors within a 30 pkpc radius of the AGNs, respectively.}
    \label{fig:agnneighbourcount}
\end{figure*}

In \S \ref{underlyingpopulation}, we delve into the statistics and properties of our underlying population of AGNs and their neighboring SMBHs, establishing a baseline for our study. Following this, \S \ref{dualmultioccurance} is dedicated to examining the occurrence rate of dual and multiple AGNs in our simulation. Then, in \S \ref{duals}, we focus on dual AGNs and conduct a detailed analysis of the characteristics of these AGNs and their host halos in comparison with those of single AGNs, highlighting differences and patterns. Finally, in \S \ref{evolution}, we trace the evolution of dual AGNs over time, exploring their developmental trajectories.

\subsection{Characteristics of the Underlying AGN Population}\label{underlyingpopulation} 

\begin{figure}
    \centering
    \includegraphics[width= 0.45\textwidth]{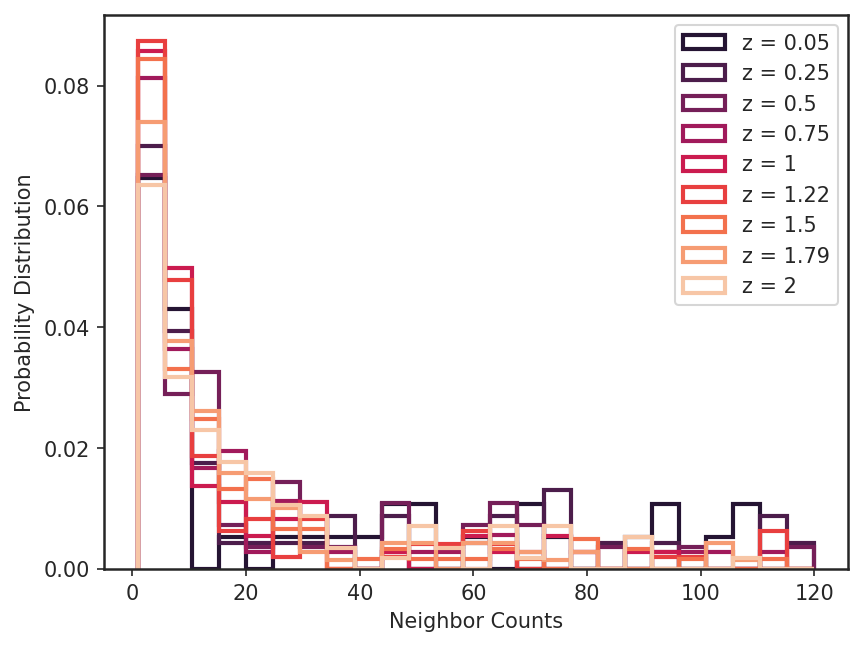}
    \includegraphics[width= 0.47\textwidth]{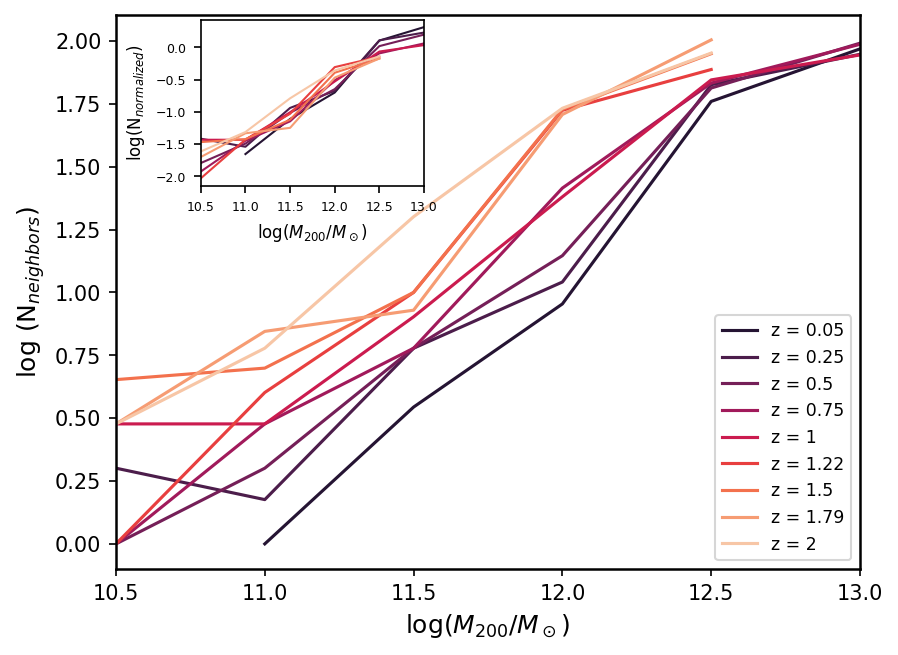}
    \caption{ The \textit{top panel} displays the probability distribution of neighbor counts around AGNs at varying redshifts normalized to integrate to unity. Lighter colors represent higher redshifts and darker colors indicate lower redshifts. The \textit{bottom panel} shows the logarithm of the number of neighbors as a function of the host AGN halo $\rm M_{200}$, with each line representing the median of neighbors counts at each mass bin. This demonstrates a near-linear relationship between halo mass and neighbor counts, with higher mass halos having more neighboring SMBHs.}
    \label{fig:neighbourcounts}
\end{figure}

We begin our analysis by examining the statistics of AGNs within the {\sc Romulus25} simulation across the redshifts under consideration. This analysis also includes their neighboring SMBHs that are situated within 30 pkpc of each AGN. As mentioned above, at this stage, no mass or luminosity thresholds are applied to these neighboring SMBHs.

The left panel of Fig. \ref{fig:agnneighbourcount} displays the number density of AGNs at each redshift, while the middle and right panels show the absolute numbers of AGNs and neighboring SMBHs, respectively. The AGN number density increases from $z$ = 0.05 to $z$ = 2, aligning with observational results of AGN number density \citep{hewitt1993, steffen2003, wolf2003, brusa2010, ceraj2018}. The AGN count rises from 40 at $z$ = 0.05 to over 140 at $z$ = 1.79, then slightly declines to 120 at $z$ = 2. The total number of neighboring SMBHs reflects a similar pattern, increasing from 1000 to more than 2600 from lower to higher redshifts.

The probability distribution of neighboring SMBH counts around an AGN at each redshift is shown in the top panel of Fig. \ref{fig:neighbourcounts}. Each distribution is normalized to integrate to unity, which facilitates comparison across different epochs. Dark-to-light colors in the figure correspond to lower to higher redshifts. At all epochs, each AGN has at least one neighbor. While most ($\sim 83\%$) of neighbor counts are below 40, they can reach up to 120, with a higher probability of encountering more than 40 neighbors at lower redshifts.

The bottom panel of Fig.\ \ref{fig:neighbourcounts} demonstrates that the number of neighbors increases with halo mass. Each line represents the median number of neighbors versus AGN host halo mass. The number of neighboring SMBHs scales roughly linearly with the halo mass, a trend stretching from dwarf galaxy halos to massive galaxy groups. These findings align with the results of previous studies on SMBH distribution in {\sc Romulus} halos. \citet{ricarte2021wanderingbhs}, who investigated all wandering SMBHs in Romulus halos, also identified a linear relationship between the number of SMBHs in a halo and their host halo mass. They showed that the majority of these wandering SMBHs originate from the centers of destroyed infalling satellite galaxies. Therefore, as higher mass halos are expected to have undergone more mergers in hierarchical assembly, having a higher number of neighbor SMBHs is not surprising.

Exploring the redshift dependency: for redshifts $z > 1$, the halo mass range is $\rm 10.5 < log(M_{200}/M_\odot) < 12.5$, while for $z < 1$, it extends to massive groups with $\rm log(M_{200}/M_\odot) \leq 13$. This expansion of the range at lower redshifts explains the observed increase in the probability of larger neighbor counts around AGNs at these redshifts.  There are more massive halos at these redshifts, and massive halos have a higher number of neighbors; thus, there is a higher count of neighbors at low redshifts. Another notable feature is that in low-mass halos, the number of neighbors decreases with decreasing redshift, reflecting the evolution of the SMBH occupation fraction towards the present. The inset subplot in the top left normalizes the median in each mass bin at each redshift by the total number of neighboring SMBHs at that redshift. This effectively removes the redshift effect, resulting in all lines converging.

\begin{figure}
    \centering
    \includegraphics[width= 0.47\textwidth]{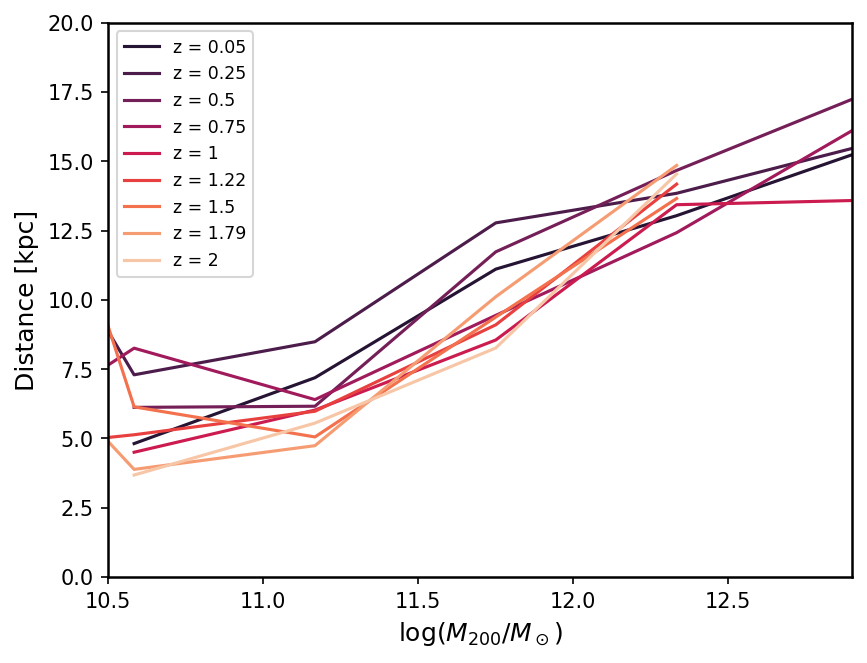}
    \caption{Relationship between halo mass and distance of neighboring SMBHs from their corresponding AGN across redshifts. Each line represents the median distance at a specific redshift. The increasing trend illustrates that higher halo masses correlate with a greater spread of neighboring SMBHs.}
    \label{fig:distancemhalo}
\end{figure}

Another insightful aspect is neighbours' spatial distribution. Fig.\ \ref{fig:distancemhalo} shows the median distance of neighbor SMBHs to their corresponding AGN as a function of AGN's host halo mass. Across all redshifts, we observe that the median distance ranges from less than 5 kpc to greater than or equal to 15 kpc. A notable trend emerges: higher halo masses correspond to larger median distances between AGNs and their neighboring SMBHs. This trend can be attributed to the higher velocity dispersion in massive halos, which extends the orbital decay time \citep{binney2008book}, causing SMBHs to be more widely dispersed. Moreover, these massive halos are more likely to experience frequent mergers, further introducing SMBHs and increasing the velocity dispersion. Consequently, larger halos not only contain a greater number of SMBHs but also their SMBHs are more dispersed.

Furthermore, we highlight that most SMBH concentrations are found within 20 kpc of the AGN. When we scale these distances by the $\rm R_\mathrm{200}$ of the AGN's host halo, we find that the majority of neighboring SMBHs reside within 0.2$\rm R_\mathrm{200}$. Those outside this boundary are typically associated with subhalos, accounting for less than 2\% of the total counts of neighboring SMBHs at each redshift.

\begin{figure*}
    \centering
    \includegraphics[width= 1.\textwidth]{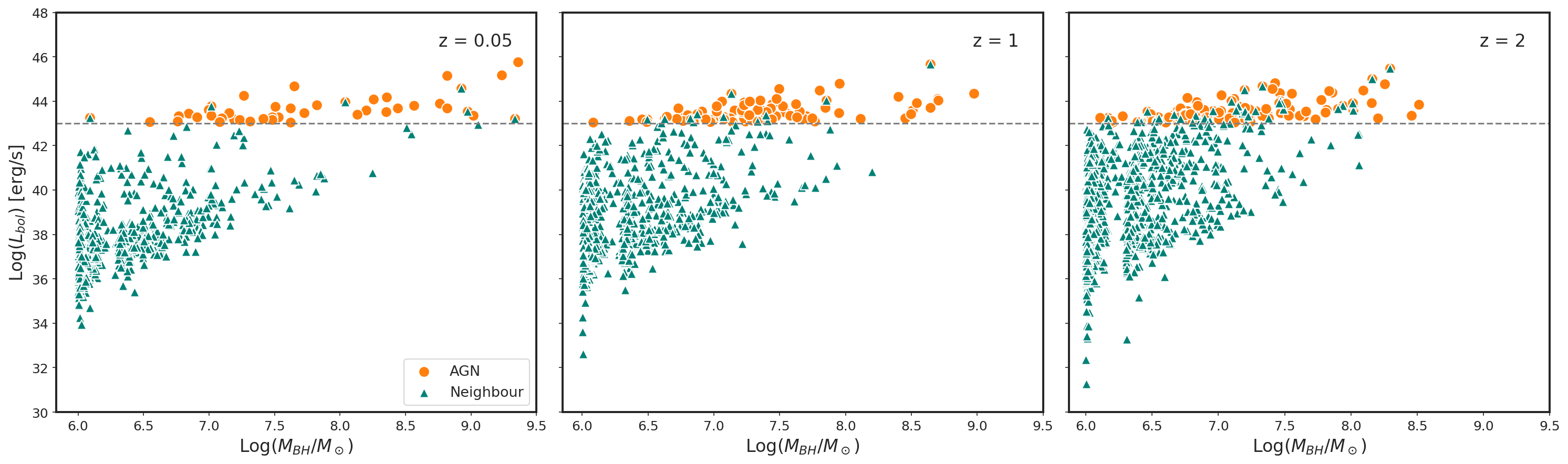}
    \caption{Bolometric luminosity as a function of SMBH mass for AGNs and their neighboring SMBH at three different redshifts (z = 0.05, 1, and 2). AGNs are represented by orange circles, while the neighboring SMBHs are depicted as teal triangles. The dashed line indicates the luminosity cut for defining AGNs.}
    \label{fig:lummass3z}
\end{figure*}

Lastly, we turn our attention to the mass and luminosity of AGNs and their neighboring SMBHs.
A variety of models predicts that neighboring SMBHs can manifest as oﬀ-nuclear X-ray or radio sources \citep{fujita2008,bellovary2010,sijacki2011,steinborn2016,barrows2019,Zivancev2020,guo2020A,bartlett2021}. In Fig.\ \ref{fig:lummass3z}, we plot the bolometric luminosities of AGNs (as orange circles) and neighboring SMBHs (as teal triangles) against their mass. The grey dashed line shows the luminosity cut used to define AGNs. 
We present illustrative results at $z$ = 0.05, 1, and 2 for brevity, but we confirm that the observed trend remains consistent across all redshifts under consideration.

At higher redshifts, we see that the SMBH mass range is generally limited to $M_\bullet < 10^8 \ \rm M_\odot$. Interestingly, even SMBHs with masses close to the seed mass ($M_\bullet \geq 10^6 \ \rm M_\odot$) contribute to high luminosity ( $L_{bol} > 10^{43} $ erg s$^{-1}$). We also note that the luminosity range for neighboring black holes can be as low as $L_{bol} \sim 10^{31} $ erg s$^{-1}$. 
As redshift decreases, the general count of AGNs with luminosity above the threshold (and also the number of neighbor SMBHs) diminishes. Their mass range expands to $M_\bullet > 10^9 \ \rm M_\odot$, and SMBHs with mass $M_\bullet < 10^7 \ \rm M_\odot$ are less likely to contribute to high luminosities. On the other hand, the minimum luminosity of neighbors increases, with their minimum be $L_{bol} \sim 10^{34} $ erg s$^{-1}$.
These findings suggest that with more sensitive instruments, a significantly larger number of AGNs with luminous companions could be detected.

\subsection{Occurrence Rates of Dual/Multiple AGNs}\label{dualmultioccurance}

\begin{figure}
    \centering
    \includegraphics[width= 0.45\textwidth]{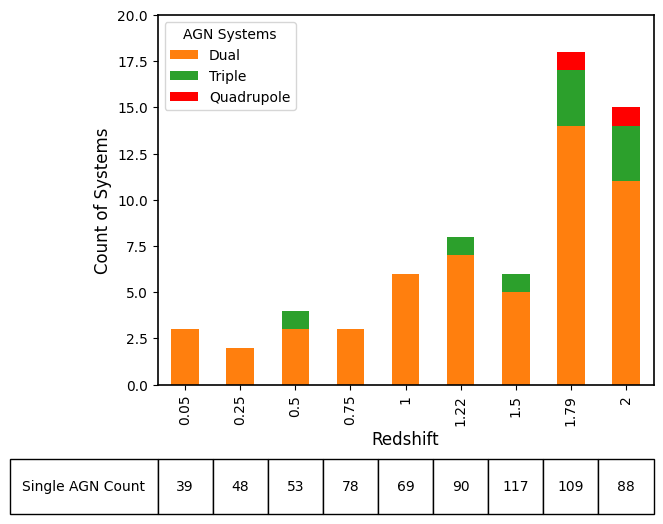}
    \caption{Count of AGN systems at different redshifts in the Romulus25 simulation, categorized by dual (orange), triple (green), and quadruple (red) AGNs. The number of single AGNs for each redshift is displayed in the table cells located below the corresponding bars in the chart.}
    \label{fig:dualmultiagnstat}
\end{figure}

\begin{figure}
    \centering
    \includegraphics[width= 0.48\textwidth]{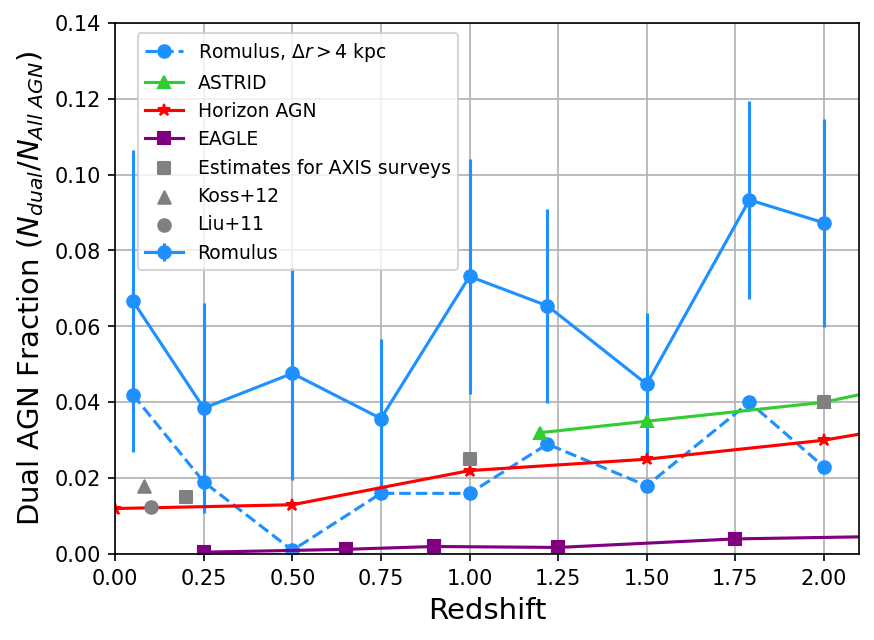}
    \caption{Predictions of the dual AGN fraction as a function of redshift from our sample in {\sc Romulus} (blue solid line) compared to other cosmological simulations: {\sc ASTRID} \citep{chen2023astrid}, {\sc Horizon-AGN} \citep{volonteri2022dualagn}, {\sc EAGLE} \citep{rosas-guevara2019} with green, red and purple lines respectively. \vida{The error bars show 1$\sigma$ standard deviation}. We also show the dual fraction in {\sc Romulus} with a separation criterion of 4 kpc < $\Delta r$ < 30 kpc with a dashed blue line. Observational estimates and predictions are presented with grey square \citep[predictions for upcoming AXIS surveys,][]{foord2023tracking}, grey triangle \citep{koss2012} and grey circle \citep{liu2011}. }
    \label{fig:dualfrac}
\end{figure}

In the Romulus25 simulations, not only dual AGNs are present but also there are instances where more than two luminous AGNs are in close proximity to each other. The occurrence of dual and multiple AGN systems, in comparison to the population of single AGNs, could serve as an indicator of the frequency of massive galaxy mergers and flag the potential binary SMBH that are nano-Hertz gravitational wave sources \citep{saeedzadeh2023bbh,Sah2024}. 

Fig.\ \ref{fig:dualmultiagnstat} presents the distribution of systems with single, dual, triple, and quadruple AGNs. It is evident that multiple (i.e. triple, and quadruple) AGNs predominantly occur at higher redshifts, with quadruple systems exclusively identified at $z$ = 1.79 and 2. Additionally, the data shows a rise in the counts of dual AGNs with increasing redshift. This trend is consistent with an increase in overall AGN number densities with redshift, as shown in Fig.\ \ref{fig:agnneighbourcount}, which correlates with a higher frequency of galaxy interactions in the higher redshifts \citep{patton2002,hopkins2007galaxyevolution}. 

We calculated the probability of an AGN having another AGN counterpart at each redshift. To do so we divided the number of AGNs within dual or multiple systems by the total AGN population. Results show that the likelihood of an AGN having \textit{at least} one AGN counterpart increases markedly from lower to higher redshifts, starting from a range of 7 to 14\% between $z$ = 0.05 and $z$ = 1.5, and jumping to 27\% at $z$ = 1.79 and $z$ = 2. 


\vida{In Fig.\ \ref{fig:dualfrac}, we focus exclusively on the fraction of dual AGNs as a function of redshift. This fraction, which is regarded as a proxy for the number of massive galaxies undergoing mergers, is calculated by dividing the number of identified dual AGN systems by the total number of AGNs \citep{liu2011,koss2012,foord2023tracking}. We do the same to
ensure comparability with observational studies. The dual AGN fraction from the {\sc Romulus} sample is shown by a solid blue line, with 1$\sigma$ standard deviation error bars assuming Poisson statistics \citep{Gehrels1986}. Looking at the {\sc Romulus} data points, it is challenging to discern a clear trend with redshift. The small box size and limited statistics cause the fraction variations with redshift to be noisy. The results can be interpreted as being consistent with a constant dual AGN fraction of $\sim$ 0.06. However, the results are also consistent with a weak rising trend. This latter interpretation aligns with findings from the {\sc ASTRID} \citep{chen2023astrid}, {\sc Horizon AGN} \citep{volonteri2022dualagn}, and {\sc EAGLE} \citep{rosas-guevara2019} simulations, which have larger box sizes ($\rm (250 \ cMpc)^3$, $ (142 \ cMpc)^3$, and $\rm (100 \ cMpc)^3$, respectively) and therefore better statistics. These find a weakly rising trend in dual AGN fractions with redshift.}

\vida{Looking at the dual fraction values,} {\sc Romulus} exhibits a higher fraction than all other simulations. This can be attributed to {\sc Romulus} having a higher spatial resolution than the simulations mentioned, allowing it to resolve dual AGNs with separations of less than 1 kpc, as well as the fact that it has a sophisticated dynamical friction model that tracks SMBH dynamics and does not force them to merge when galaxies merge. Among other simulations, {\sc EAGLE}
\citep{rosas-guevara2019} does not implement a dynamical friction model for SMBHs and reports dual AGNs with a minimum separation of 5 kpc. {\sc Horizon AGN} includes a dynamical friction model, but its merging criteria merge SMBHs with separations of less than 4 kpc \citep{volonteri2022dualagn}, thus not allowing for the resolution of duals below that separation. {\sc ASTRID} includes a dynamical friction model and also has a higher resolution than both {\sc EAGLE} and {\sc Horizon AGN}, with a spatial resolution of $\sim$ 1.4 kpc, and exhibits a higher fraction than the others \citep{chen2023astrid} as well. The observational results also suffer from the inability to resolve closely separated dual AGNs. All mentioned observational estimates \citep{liu2011,koss2012} as well as the theoretical estimates for upcomming AXIS surveys \citep{foord2023tracking} report \vida{results for AGN} separations of greater than 5 kpc. It's worth noting that if we limit our sample to duals with separations greater than 4 kpc, our dual AGN fraction (shown by a dashed blue line) decreases and agrees with other simulations and observational estimates. \vida{In our sample, 65\% of dual AGNs have separations of 0.5 - 4 kpc, while only 35\% are separated by 4 - 30 kpc.}
These results predict a substantial number of dual AGNs with small separation; motivating deep surveys and sensitive instruments capable of testing this prediction.


\subsection{Properties of Dual AGNs} \label{duals}

Having explored the occurrence rates of dual and multiple AGNs, we now narrow our focus to dual AGNs for a more detailed analysis. We only concentrate on dual AGNs because of the relatively low statistical representation of multiple AGNs in our simulations, which limits the robustness of any analysis we could conduct on them. Here we explore dual AGNs properties, such as their separation, luminosity, mass, and the characteristics of their host halos.

\vspace{1cm}

\subsubsection{Separation}

\begin{figure}
    \centering
    \includegraphics[width= 0.45\textwidth]{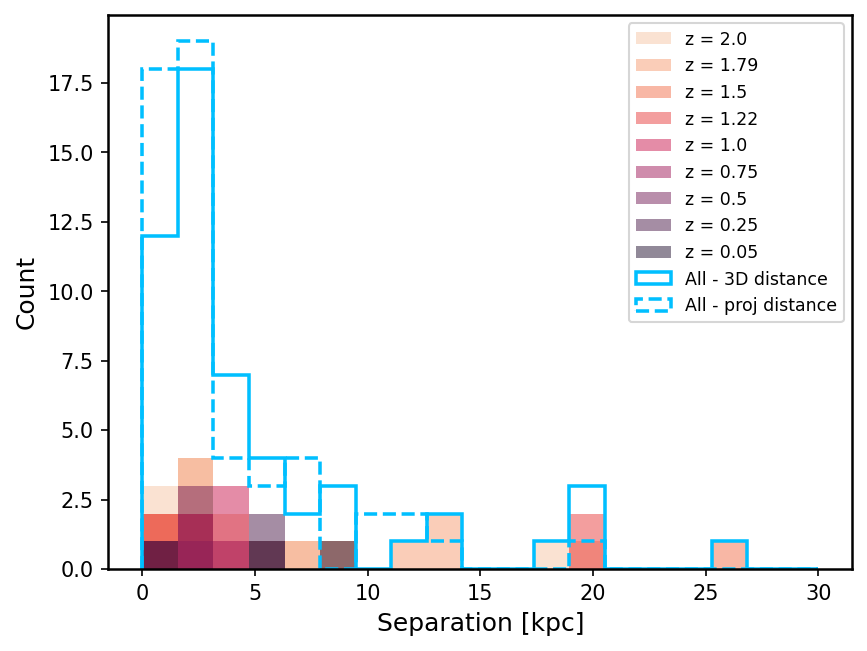}
    \caption{ Distribution of separations between SMBHs in dual AGNs at redshifts under consideration. Each redshift is represented by a distinct color, with filled histograms indicating the count of dual AGNs at that specific redshift. The solid blue line outlines the combined count for all redshifts, illustrating the 3D physical separation between dual AGNs, while the dashed blue line represents the projected 2D separations on the x-y plane. The plot highlight a higher frequency of dual AGNs with separations less than 10 kpc, peaking around 2 kpc.}
    \label{fig:dualseparation}
\end{figure}

Fig.\ \ref{fig:dualseparation} illustrates the distribution of separations between the two SMBHs in dual AGNs across the redshifts under consideration. The distribution at each redshift is represented by filled histograms, while the combined data for all redshifts is shown in an unfilled blue histogram. The solid blue line indicates the 3D physical separation. The projected physical separation is also shown by a dashed blue line to mimic the selection function of observations. Here, we take the projected separation to be the projection of the 3D separation onto the x $-$ y plane.

The blue solid histogram indicates that the majority of dual AGNs in our sample have separations below 10 kpc, with a notable increase within 5 kpc and a peak around 2 kpc. In our sample, the probability of observing a dual AGN with a separation of less than 2 kpc is seven times higher than at larger separations. The 2D projected separation histogram shows closer separations between the duals, resulting in an increase in duals with separation < 1 kpc and a maximum separation of 20 kpc (compared to 27 kpc in 3D). The overall trend is consistent with the 3D data, showing a higher concentration of duals within 10 kpc and a peak at 2 kpc. Since our simulation adopts a sub-grid dynamical friction model and it has a high resolution, it allows for the identification of dual AGNs at separation $\sim$ 0.5 kpc. In previous works, \citet{rosas-guevara2019} observed a peak in dual separations near $20 - 25$ kpc, but their study excluded pairs below 5 kpc. Other simulations using different sub-grid dynamical friction models which allow black holes to approach closer than 5 kpc before merging, have similarly found a higher probability density of duals at separations less than 5 kpc, aligning with our findings \citep{steinborn2016,Volonteri2022dualhorizon,chen2023astrid}.

In our comparative analysis of the dual AGN population across various redshifts, we ﬁnd interesting diﬀerences. Specifically, we find that at $z$ < 1, all dual AGN in our sample shows a separation of less than 10 kpc. At higher redshifts (z > 1), the number of dual AGN systems with separations higher than 10 kpc increases, such that 23\% of dual AGNs at these redshifts have separations exceeding 10 kpc.  This variance suggests an evolution in the spatial dynamics of dual AGNs over cosmic time. All duals are detected to be in the same halo at the time of observation, where the host halos are identiﬁed with the subhalo catalog generated by AHF. The only exception is the SMBHs in a dual AGN with a 27 kpc separation at $z$ = 1.5, which are located in two different halos.

\subsubsection{Mass and Luminosity}

\begin{figure}
    \centering
    \includegraphics[width= 0.45\textwidth]{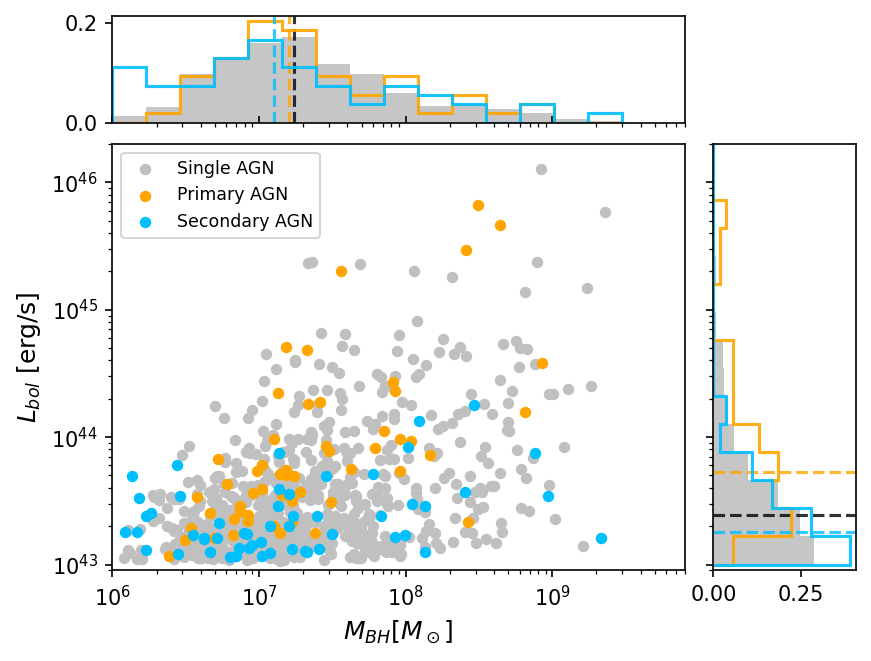}
    \caption{The relationship between the bolometric luminosity and mass of SMBHs in single and dual AGN systems. Single AGNs are represented by grey dots, primary AGNs in dual systems by orange dots, and secondary AGNs by blue dots. The top panel shows the normalized distribution of SMBH mass for the different populations, while the right panel displays the normalized distribution of bolometric luminosity. Dashed lines showing the medians.}
    \label{fig:duallummass}
\end{figure}

\begin{figure}
    \centering
    \includegraphics[width= 0.45\textwidth]{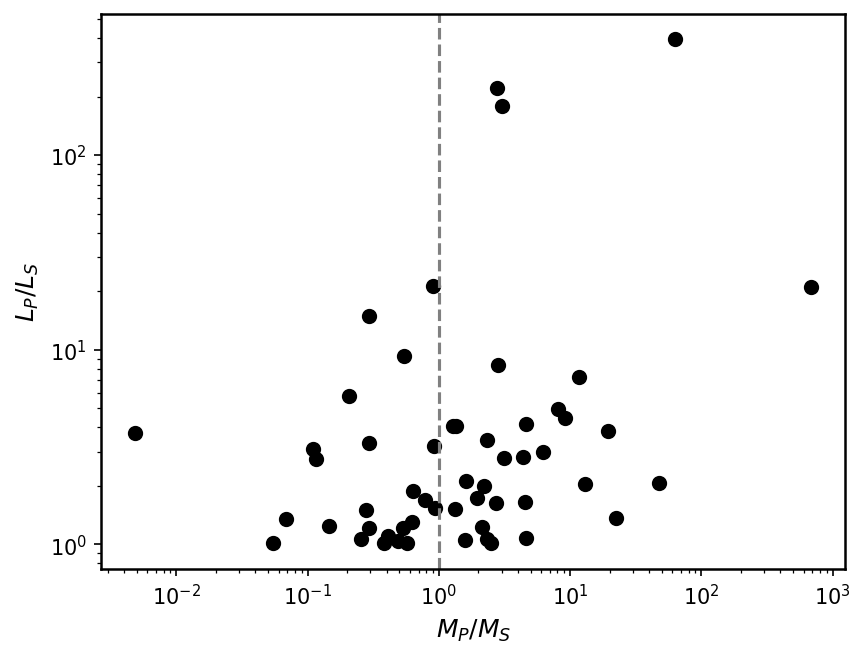}
    \caption{ Scatter plot illustrating the ratio of bolometric luminosities to the mass ratio of primary to secondary supermassive black holes (SMBHs) in dual AGN systems. Each point represents a dual AGN, plotted with respect to the mass ratio on the x-axis and luminosity ratio on the y-axis. The dashed vertical line at $M_p/M_s = 1$ demarcates equal mass pairs. }
    \label{fig:duallummassfrac}
\end{figure}

Fig.\ \ref{fig:duallummass} displays the relationship between luminosity and mass for the two SMBHs involved in the dual AGN systems at all redshifts under consideration, alongside the population of single AGNs for comparison.
A single AGN is an AGN that none of its neighboring SMBH has luminosity above the threshold of $10^{43} $ erg s$^{-1}$. 
The central panel shows the scatter plot of luminosity against SMBH mass, while the upper and right panels present the 1D distributions of SMBH masses and luminosities, respectively. In this figure, the primary (brighter) SMBHs of the dual AGNs are represented in orange, while the secondary (fainter) SMBHs are indicated in blue. \vida{Both AGNs necessarily have luminosity > $10^{43} $ erg s$^{-1}$}. The single AGNs are plotted in grey. Each histogram is normalized to the total population of SMBHs within the respective category they represent and dashed lines show the median of the properties for each category.

The luminosity of the AGNs ranges from our threshold of $10^{43}$ to $ 10^{46} $ erg s$^{-1}$and masses range from the simulation SMBH seed mass of $10^6 $ to $ 3 \times 10^9 \rm \, M_\odot$. From the scatter plot and histograms, we find that the ranges of luminosity and mass for dual AGNs are not significantly different from those of single AGNs. This finding is in agreement with the observational results reported by \citet{tang2021silverman}.

Examining the top panel of Fig.\ \ref{fig:duallummass}, we see that the shape of the mass histograms for the primary and secondary SMBHs in dual AGNs, as well as for single AGNs, are generally similar, each peaking at approximately $\rm M_{BH} \sim 10^7 M_\odot$. The main differences compared to single AGNs are: (i) a higher proportion of primary SMBHs in dual AGNs have a mass of $10^7 M_\odot$ and (ii) secondary SMBHs in dual systems more frequently exhibit masses lower than $10^7 M_\odot$. This suggests that a low-mass black hole with a bolometric luminosity greater than  $L_{bol} > 10^{43} $ erg s$^{-1}$ is more likely to be part of a dual system than to be a single AGN.

In terms of luminosity distribution (right panel, Fig.\ \ref{fig:duallummass}), the secondary SMBHs exhibit a pattern similar to that of single AGNs, with a large proportion clustered around the threshold $L_{bol} \sim 10^{43} $ erg s$^{-1}$. Conversely, the luminosity distribution of primary SMBHs in dual systems skews towards higher values, a trend that arises from our definition of the primary SMBH as the more luminous of the pair. Notably, a greater fraction of primary SMBHs display luminosities exceeding $ 10^{44} $ erg s$^{-1}$ compared to single AGNs. 

\vida{To further investigate the relationship between luminosity and dual AGN systems, we examined the fraction of AGNs that are part of a dual system above specific luminosity thresholds. We calculated this fraction by dividing the number of AGNs that are part of a dual AGN system and have a luminosity above a specific threshold by the total number of AGNs above that luminosity threshold. We used $L_{bol} = 10^{43}, 10^{44}$, and $10^{45} $ erg s$^{-1}$ as thresholds. The results show that of the AGNs with luminosities above $10^{43}$, $10^{44}$, and $10^{45} $ erg s$^{-1}$, 13\%, 14\%, and 29\% respectively are part of a dual AGN. This indicates that the probability of an AGN being part of a dual system increases with increasing luminosity. The significant fraction of the brightest AGNs being part of a dual AGN system supports findings that suggest galaxy mergers enhance AGN activity during merger and post-merger phases \citep{Ellison2019,ByrneMamahit2023}.}

\vida{We note that the above values were computed using AGNs across all redshifts in our sample. To make meaningful comparisons with observational studies, it is essential to apply the same selection criteria used in those studies. For example, applying the same criteria as \citet{liu2011} \footnote{To compare our results with \citet{liu2011}, we considered AGNs with $L_{bol} > 10^{43}$ erg s$^{-1}$, separations of 5 kpc < r < 30 kpc, and z < 0.16.} the fraction of AGNs in dual systems drops to 4.4\%. However, before comparing this result to the 1.3\% reported by \citet{liu2011}, one more consideration must be addressed: \citet{liu2011} report the fraction based on the number of AGN pairs. After accounting for this, our corresponding result is 2.2\%.}

We explore the primary to secondary SMBHs luminosity and mass ratio in Fig.\ \ref{fig:duallummassfrac}. The luminosity ratio can exceed values of 10, though only about 10\% of duals present such a high luminosity contrast, with the primary AGN being more than ten times brighter than the secondary. This high contrast has been reported in observational works such as \citet{koss2012} and has been seen in various simulations \citep[e.g.][]{callegari2009,steinborn2016,capelo2017,chen2023astrid} as well. The majority of duals in our sample display a more moderate average luminosity contrast of 3 to 1.

Notably, there are 45\% of duals with mass ratio $M_p/M_s$ < 1 and luminosity ratio $L_p/L_s$ > 1, indicating that the lower mass SMBH in the dual AGN system is outshining the more massive one. We investigate the environment resulting in such cases in the next section (\S \ref{evolution}). Our findings suggest that the local gas density within the vicinity (less than 1 kpc) of SMBHs plays a crucial role in instances when the less massive SMBH becomes more luminous than its more massive counterpart. Specifically, a higher gas density surrounding the less massive black hole leads to increased accretion rates, thereby boosting its luminosity. This is predominantly observed when the less massive SMBH is near pericenter and gas fueling and the loss of angular momentum are enhanced.  This result is in agreement with idealized high-resolution binary black hole merger simulations \citep[e.g.][]{capelo2017}

\subsubsection{Host halos}

\begin{figure}
    \centering
    \includegraphics[width= 0.45\textwidth]{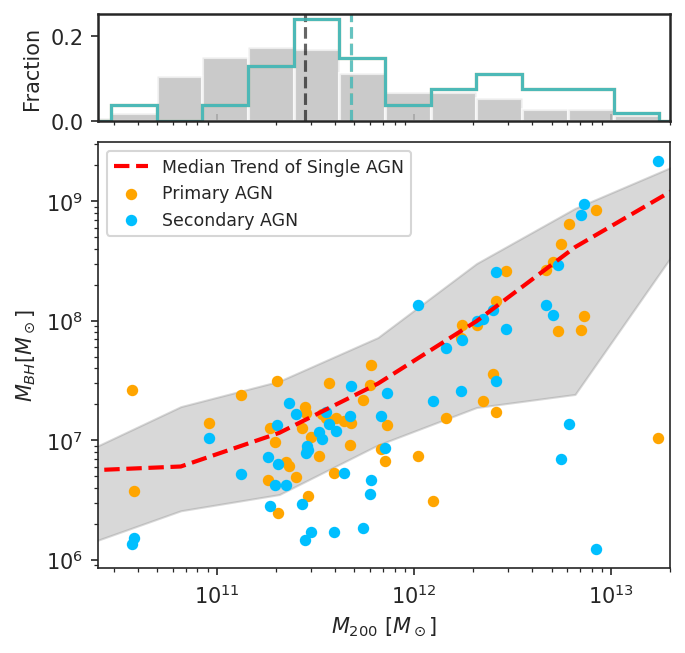}
    \caption{Relationship between Black Hole Mass and Halo Mass for Dual AGNs. This scatter plot illustrates the primary AGN masses in orange and the secondary AGN masses in blue, plotted against their host halo mass. The dashed red line indicates the median trend of single AGNs for comparison. The shaded region encloses the scatter of the 16 to 84\% of the MBH in the halo mass bin. The histogram at the top displays the normalized count of the halo mass distribution for single AGN hosts in grey and for dual AGN host halos in cyan. Dashed lines show medians.}
    \label{fig:bhmasshalo200}
\end{figure}

\begin{figure*}
    \centering
    \includegraphics[width= 1.\textwidth]{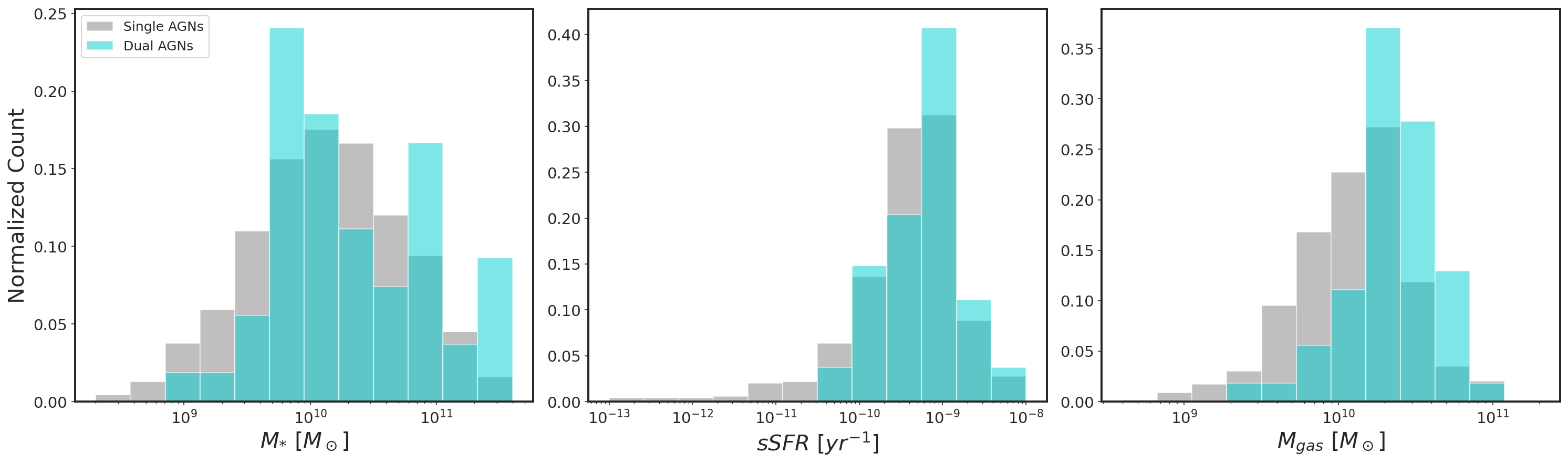}
    \caption{Comparative Distribution of AGN Host Properties. The left panel displays the stellar mass distribution, the center panel shows the specific star formation rate gas mass distribution and the right panel illustrates the gas mass distribution for single (grey) and dual (cyan) AGNs inside a 30 kpc sphere around their host halo center.}
    \label{fig:mstarmgasssfr}
\end{figure*}

In this section, we shift our focus to the host halos of dual AGNs, comparing their properties with those of single AGN hosts. Fig.\ \ref{fig:bhmasshalo200} central panel illustrates the $\rm M_{BH} - M_{200}$ relationship for dual and single AGNs and their respective host halos, while the top panel presents the one-dimensional distribution of the halo masses with histograms and their median with dashed lines. The mass distributions of all single AGNs are shown in grey. The red dashed curve represents the median $M_{BH}$ for these AGNs within each halo mass bin and the shaded region encloses the scatter of the 16 - 84\% of the $M_{BH}$ in that halo mass bin. The masses of the primary and secondary AGNs are represented in orange and blue, respectively, plotted against the mass of their host halo. Our sample included only one dual AGN system in separate halos, which is excluded from this analysis to maintain consistency, as all other dual systems are in the same halo.

From the 1D distribution of halo mass, we see that dual AGNs, shown with cyan histogram, favor the more massive halos. For single AGNs, the halo masses peak at $ 2 \times 10^{11} M_\odot$. 
The host halo mass distribution for dual AGNs exhibits a bimodal pattern, with the first peak being only marginally higher than that for single AGNs, at $3 \times 10^{11} M_{\odot}$. The second peak occurs at $3 \times 10^{12} M_{\odot}$. Notably, the occurrence of dual AGNs significantly decreases in less massive halos.
Furthermore, a larger proportion of dual AGNs' host halos exhibit $M_{200} > 10^{12} M_{\odot}$. Part of the reason for the skewed halo mass distribution of duals is that the dual AGN mostly selects out primaries that are on the massive end of the single AGN population (see Fig.\ \ref{fig:duallummass}). Comparing the $M_{BH} - M_{200}$ relation of duals with that of single AGNs in the central panel, we see that more than 80\% of pairs have halo masses above the median relation in the same SMBH mass bin of single AGNs, meaning that SMBHs in duals are under-massive relative to their host halos. This finding aligns with results from a previous study by \citet{steinborn2016} on $z$ = 2 AGN pairs, which noted that the $M_{BH}$ in these pairs was systematically under-massive compared to their host masses. One possible explanation is that the host halos of SMBHs that are now involved in a dual AGN system merge but it takes a longer time for the SMBHs themselves to merge and meet the median $M_{BH} - M_{200}$ relation of single AGNs. We tested this hypothesis by comparing the sum of the SMBH masses in each dual system against halo mass; the combined SMBH masses fall within the grey shaded area that matches the single AGN mass-halo mass relationship.

Fig.\ \ref{fig:mstarmgasssfr} compares the stellar mass, specific star formation rate (sSFR) and gas mass within a 30 kpc radius from the centers of host halos for dual AGNs with those of single AGN hosts. In our analysis, we normalize the histogram counts for each property by the total counts of single and dual AGNs, respectively. The first panel, which focuses on stellar mass, reveals similar distributions for both dual and single AGN host galaxies. This finding aligns with the observation of \citet{stemo2021}, who reported no significant differences in the galaxy mass distributions between AGN pairs and single AGN samples. Our results, however, appear to contrast with those of \citet{chen2023astrid}, who found dual AGNs residing in significantly larger mass galaxies compared to single AGNs, the discrepancy primarily stems from the differences in how dual and single AGNs are defined in our study versus that of \citet{chen2023astrid}. \citet{chen2023astrid} define SMBHs with mass greater than $10^7 M_\odot$ as single AGNs, whereas we classify SMBHs with a $L_{bol > }10^{43} $ erg s$^{-1}$as single AGNs. When applying a luminosity threshold to define single AGNs, \citet{chen2023astrid} also find results indicating no significant differences between hosts of the dual and single AGN populations.

Furthermore the first panel of Fig.\ \ref{fig:mstarmgasssfr} shows that dual AGNs are less commonly found in lower-mass galaxies, specifically those with a stellar mass less than $5 \times 10^9 M_\odot$. This suggests that hosting two sufficiently luminous AGNs typically requires a galaxy to have undergone at least one relatively major merger, and that massive galaxies experience, overall, more mergers than their lighter counterpart \citep[e.g.][]{rodriguez-gomez2015,dubois2016}. Additionally, a dichotomy is seen in the dual AGN distribution, with peaks at $10^{10} M_\odot$ and $7 \times 10^{10} M_\odot$. 

In the second panel of Fig.\ \ref{fig:mstarmgasssfr}, we show sSFR distribution. To calculate sSFR, we sum the SFR within a 30 kpc radius from the host halo center and then divide this by the total stellar mass within the same radius. Both dual and single AGNs show similar sSFR distribution shapes, with a central value around $10^{-9} yr^{-1}$. However, dual AGNs are shifting more 
towards slightly higher sSFR values and are not represented as much in the low sSFR tail, which is noticeable in single AGN host halos. This trend aligns with findings from previous studies using idealized galaxy merger simulations \citep[e.g.][]{van-wassenhove2012}, which found peaks in the host galaxies’ SFR after several pericentric passages when the duals are separated by only a few kpc.

Lastly, the distribution of the gas mass for dual AGNs exhibits a peak similar to that of single AGNs, yet it is characterized by significantly higher fractions at higher  $M_{gas}$ values. This pattern suggests that dual AGNs are more likely to be found in halos with higher gas mass compared to the average single AGN host halos.


\subsection{Evolution of Dual AGN} \label{evolution}

\begin{figure}
    \centering
    \includegraphics[width= 0.45\textwidth]{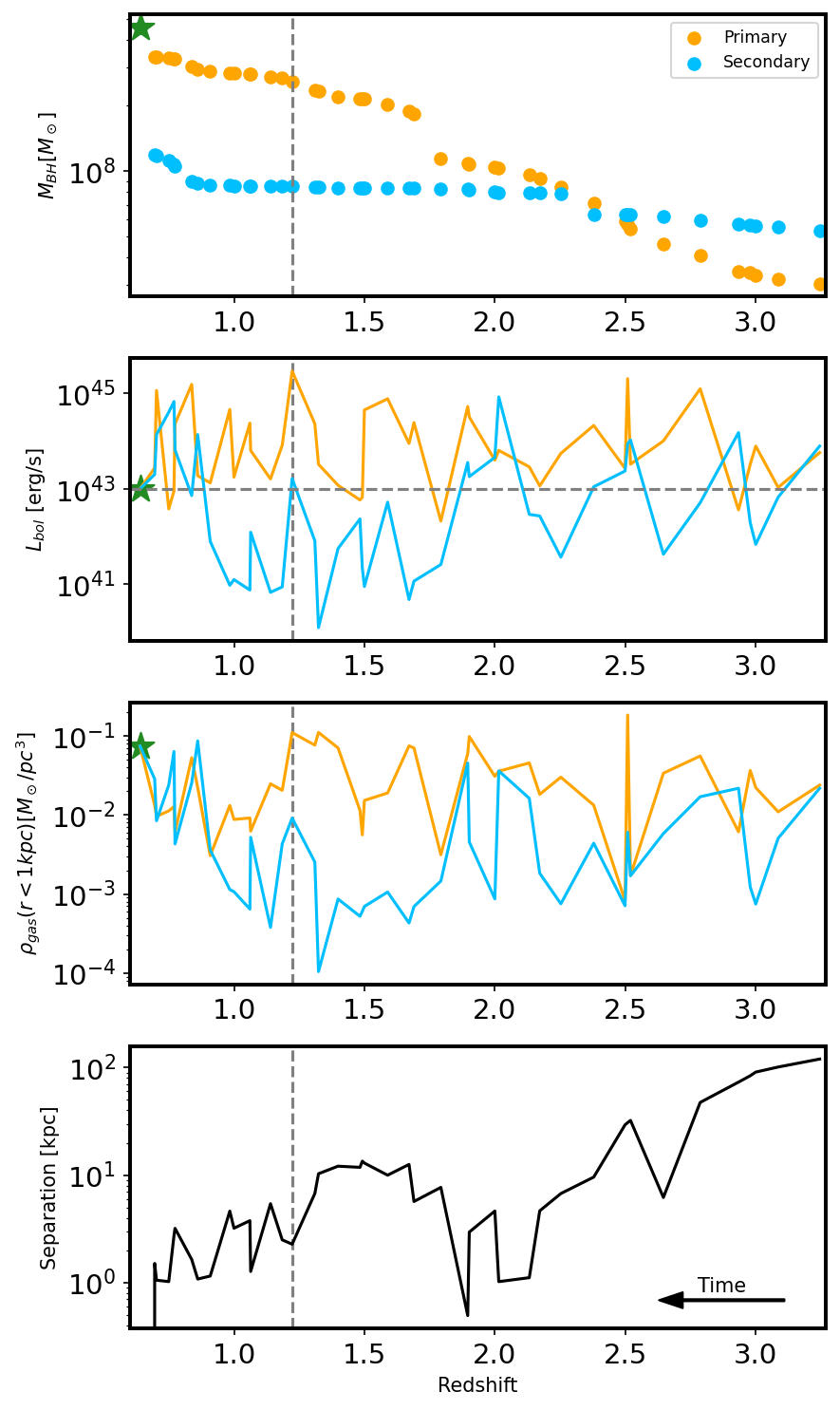}
    \caption{Evolution of dual AGNs that merge. The figure shows the evolutionary properties of a dual AGN system from the time that they were in two distinct halos until they merged (shown by a green star). We trace their masses (first panel), luminosity (second panel), surrounding gas density (third panel), and pair separation (fourth panel). The dashed vertical line shows the redshift that duals have been detected. We remind the reader that, in our study, detection is confined to selected epochs at $z$ $\leq$ 2 (see Section \ref{sec:method-selection}). The dashed horizontal line is the luminosity threshold for AGN definition in our study. }
    \label{fig:evolutionmerged}
\end{figure}


\begin{figure}
    \centering
    \includegraphics[width= 0.45\textwidth]{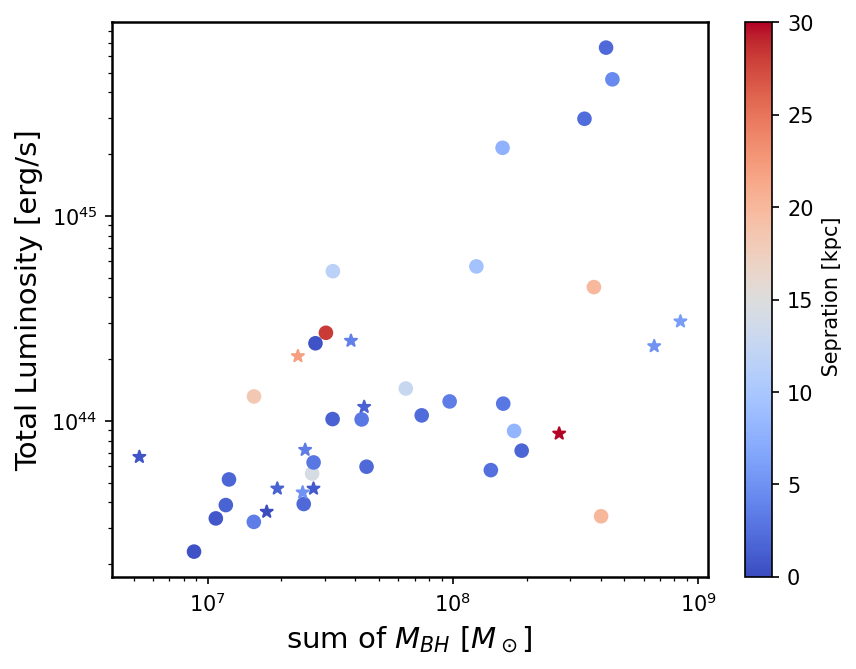}
    \includegraphics[width= 0.45\textwidth]{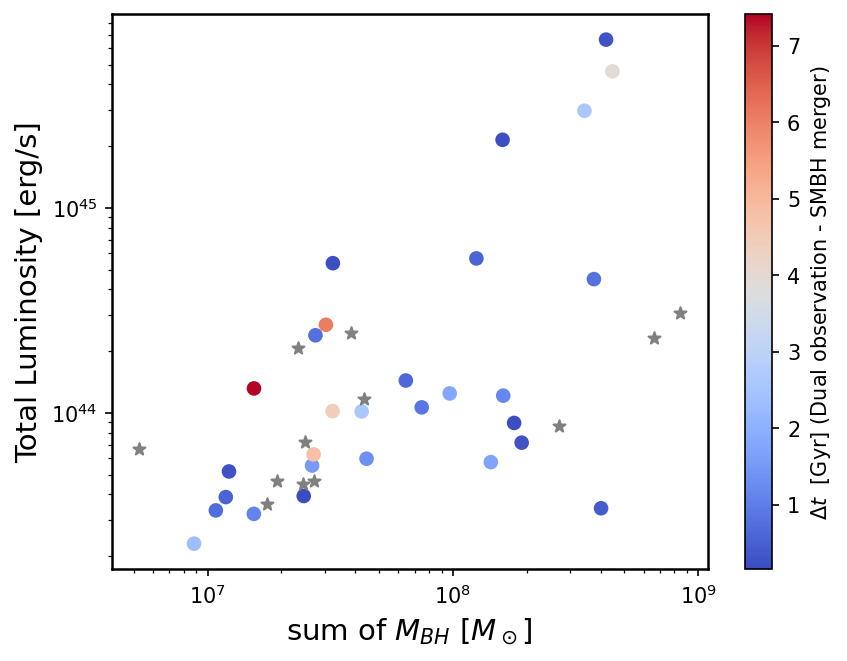}
    \caption{ The total luminosity of two SMBHs in a dual AGN system versus their combined mass. SMBHs within a dual system that eventually merge are represented with circles, while those that do not merge by $z$ = 0 are represented by stars. In the \textit{top panel}, the data points are color-coded based on the separation between the two SMBHs when they are detected as a dual AGN. The data points in the \textit{bottom panel} are color-coded according to the time elapsed between the detection of the dual AGN and the merger of the two SMBH. The stars are shown in grey because the merging time is not applicable to them.}
    \label{fig:mergervsnonmerger}
\end{figure}

\begin{figure}
    \centering
    \includegraphics[width= 0.45\textwidth]{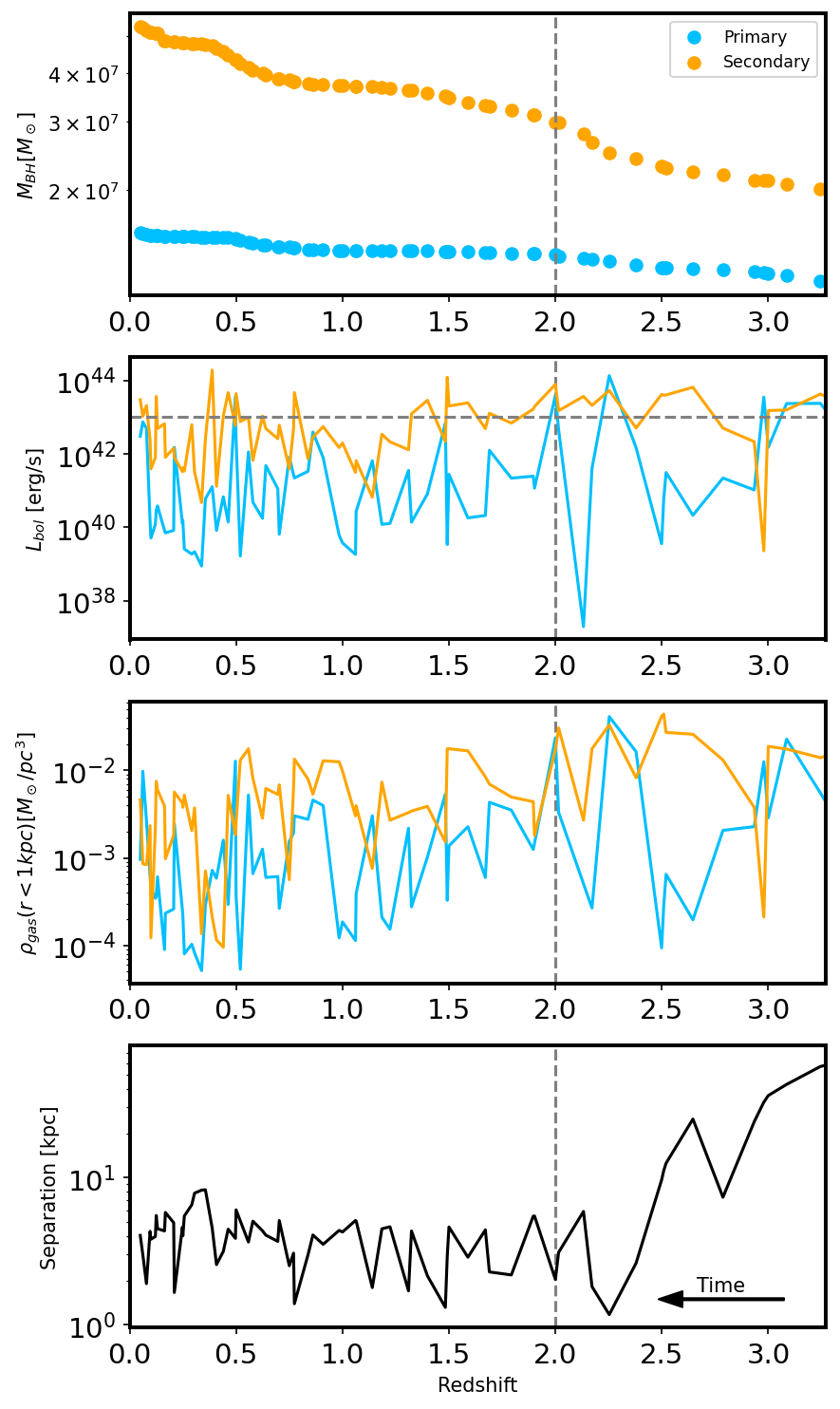}
    \caption{Evolution of dual AGNs that do not merge. The figure shows the evolutionary properties of a dual AGN system from the time that they were in two distinct halos until $z$ = 0. The colors and properties of each panel are the same as Fig.\ \ref{fig:evolutionmerged}.}
    \label{fig:evolutionnotmerge}
\end{figure}

To deepen our understanding of the evolution of dual AGNs and their relation to galaxy and SMBH mergers, we leverage the simulation's capacity to trace the evolution of these pairs across time. Specifically, we follow the SMBHs within dual AGN systems forward in time until they either merge or reach $z$ = 0. For this analysis, we consider the SMBHs to be merged when they are numerically merged in the simulation (see \S \ref{sec:romulus-bhmerger} for SMBH merging criteria). Additionally, we trace their history back to the point where they resided in separate halos, with their host halos completely distinct (i.e., the host halos are outside of each other's $\rm R_\mathrm{200}$). Note that all \footnote{There is only one dual AGN in our sample that the SMBHs are in two different halos at the time of analysis.} dual AGNs we are analyzing here are in the same halo at the time of analysis. Based on our analyses, we find that our dual AGN sample includes rapidly evolving systems, slower ones, and even ineffective SMBH mergers. We show illustrative case studies in Figs. \ref{fig:evolutionmerged} \& \ref{fig:evolutionnotmerge}.

Fig. \ref{fig:evolutionmerged} presents the evolutionary trajectory of dual AGNs that are detected at $z$ = 1.22 (marked by a vertical dashed grey line). The chosen dual AGN is typical among our sample, although there is a large variance among the population. From top to bottom, the panels display the SMBH mass, bolometric luminosity, gas density, and AGN separation as a function of redshift. The gas density is computed within a 1 kpc sphere around each SMBH. The primary and secondary SMBH, defined as the brighter and fainter AGN \textit{at the time of analysis}, are represented in orange and blue, respectively. 

In this system, the initially less massive SMBH is situated in an environment characterized by high gas density. During the halo mergers, this SMBH undergoes rapid growth via accretion and, by the time of analysis, becomes the more massive and brighter AGN. Its luminosity, though varying over time, mainly remains above $10^{43} $ erg s$^{-1}$.

The secondary SMBH experiences more gradual growth in mass. Its luminosity is on average below $10^{43} $ erg s$^{-1}$with occasional peaks above it. The peaks correlate with changes in the surrounding gas density. As the gas density increases, so does the AGN's luminosity. This pattern results in multiple epochs where the dual AGN is detectable, although these periods are typically short-lived.

\begin{table}[htbp]
\centering
\caption{Merger and Non-Merger Rates of dual AGNs at Different Separations}
\begin{tabular}{ccc}
\toprule
Separation & Merger Rate & Non-Merger Rate \\
\midrule
< 5 kpc  & 73\% & 27\% \\
> 5 kpc  & 66\% & 34\% \\
< 10 kpc & 68\% & 32\% \\
> 10 kpc & 78\% & 22\% \\
< 15 kpc & 71.5\% & 28.5\% \\
> 15 kpc & 66\% & 33\% \\
\bottomrule
\label{table:mergervsnonmerger}
\end{tabular}
\end{table}

After its second pericentric, the gas surrounding the secondary SMBH becomes disrupted, leading to a decrease in its luminosity. This trend continues until the secondary SMBH approaches its next pericentric passage, orbiting very close around the primary black hole, where both become highly active.
Eventually, the two SMBHs become very close and merge, forming a final SMBH with a mass of $5 \times 10^8 M_\odot$. 

The case study above provides an illustrative example of a dual AGN in which the SMBHs eventually merge. However, as mentioned at the beginning of this section, not all SMBHs in a dual AGN system end up merging. This is not surprising. Galaxy mergers do not invariably lead to SMBH mergers, as SMBH dynamics can be inefficient on both large \citep{dosopoulou2017,tremmel2018dancing,pfister2019,li2020,bortolas2020,bortolas2021} and small scales \citep{begelman1980,milosavljevi2001,munoz2019}.

Among SMBHs in dual AGNs systems in our sample, 70\% will end up merging \footnote{In this analysis of linking dual AGN to SMBH mergers we do not include
dual AGN at $z$ = 0.05, since this is the last redshift we have SMBH information in the simulation, and by deﬁnition they cannot give rise to an SMBH merger by the same redshift.} , 25\% stay in close orbit as stalled pairs and the remaining 5\% involve cases where one of the black holes in the duals is ejected from the system due to another galaxy merger. The fraction of successful mergers reaches 87\% for duals powered by SMBHs both with mass $> 3 \times 10^7 M_\odot$. This fraction also increases, for duals surrounded by stellar density higher than 0.24 $M_\odot/pc^3$. This is because the simulation includes star dynamical friction.

In Fig.\ \ref{fig:mergervsnonmerger}, we present the total luminosity of dual AGNs versus the sum of the masses of SMBHs in these dual systems. Circles represent duals whose SMBHs merge, while stars indicate non-merging ones. In the top panel, the data points are color-coded by separation. In the bottom panel, points are color-coded by the time span between the detection of dual AGNs (i.e., when both AGNs have luminosities above our threshold of $10^{43}$ erg s$^{-1}$ within the considered redshifts) and the subsequent SMBH mergers, for cases where a merger occurs. Stars are shown in grey because the merging time difference is not applicable to them. Note that in our study, the detection redshift range is restricted to $z$ $\leq$ 2.

As previously mentioned, 70\% of the black holes in dual AGNs are projected to merge. If we limit the sample exclusively to dual AGNs with a separation of less than 5 kpc, the fraction expected to merge slightly increases to 73\%, with 27\% not merging. Conversely, for AGNs with a separation greater than 5 kpc, 66\% are expected to merge, and 34\% are not expected to merge. Table \ref{table:mergervsnonmerger} shows the fractions with 10 and 25 kpc separation cutoffs. The results show that the merging fraction is only marginally sensitive to the separation cut-off, implying that both merging and non-merging duals are found across all separations at similar rates.

The top panel of Fig. \ref{fig:mergervsnonmerger} displays a wide range of timescales between the detection of dual AGN and SMBH mergers, spanning from a few hundred Myr to 7 Gyr. Most duals merge in less than 5 Gyr. It is only for those with a separation greater than 15 kpc that the timescale exceeds 5 Gyr. However, this is not an absolute rule; there are also duals with separations up to 20 kpc that merge within a timescale of less than 2 Gyr. 
The total luminosity of these duals ranges between $2 \times 10^{43}$ and $8 \times 10^{45}$  erg s$^{-1}$, and their total mass is  $M_{BH} > 5 \times 10^7 M_\odot$. Fig. \ref{fig:mergervsnonmerger} highlights that all duals with a total bolometric luminosity $L_{bol} > 3 \times 10^{44}$ erg s$^{-1}$will merge, regardless of their separation, which can be as large as 15 kpc. These duals will merge within a timeframe of less than 4 Gyrs.

We note that what we refer to as a `merger' is a `numerical merger in the simulation'. In Romulus, two SMBHs are considered to merge when they are closer than 0.7 kpc and gravitationally bound to each other. Therefore, what we actually observe can be described as a `successful pairing'. This is because, at scales below tens of parsecs, numerous bottlenecks can slow down or suppress SMBH binary formation and further decay \citep{deRosa2019review,amaro-seoane2023}. For the most massive and luminous SMBH systems, specifically those above $10^8 M_\odot$, our simulation's resolution is sufficiently close to the separation at which binary formation occurs. This proximity makes the extrapolation to actual mergers relatively reliable. This assertion is supported by the findings of \citet[][in preparation]{mayer2024merger}, which indicate that massive SMBHs are the most likely candidates for actual mergers.


Figure \ref{fig:evolutionnotmerge} presents an illustrative example of a dual AGN system, where the SMBH pair persists without merging until $z$ = 0. Initially, both black holes have masses within an order of magnitude of the seed mass ($10^6 M_\odot$). As the system evolves, both SMBHs experience slow growth primarily through accretion, with intermittent AGN activity. The higher-mass SMBH consistently maintains a luminosity above our defined AGN threshold until $z$ = 1.5, whereas the lower-mass SMBH only occasionally exceeds this threshold. These luminosity peaks are aligned with its pericentric passages, coinciding with times when it encounters regions of higher gas density.

At the redshift of analysis, the secondary SMBH is in its third pericentric passage, causing its luminosity to surge above $10^{43} $ erg s$^{-1}$. Following this, it settles into a steady orbit around the primary SMBH at separation < 5 kpc until $z$ = 0, during which its surrounding gas density, and consequently its luminosity, gradually decreases. Sometimes, the luminosity of the secondary drops to as low as $L_{bol} < 10^{40} $ erg s$^{-1}$. Meanwhile, The primary SMBH's luminosity stays high for $\sim$ 1 Gyr after the time of analysis and then decreases as its surrounding gas density decreases as well. These observations indicate that when two high-accreting SMBHs maintain close orbits over an extended period, they eventually deplete the gas in their vicinity, affecting the AGN activity of both. 

This analysis underscores the importance of local environmental properties, particularly gas density, in influencing AGN activity. A notable manifestation of this is when a less massive black hole outshines its larger counterpart. A more detailed analysis reveals that when the less massive SMBH goes through a region with higher gas density than that surrounding the more massive SMBH, its luminosity tends to increase, potentially surpassing that of the larger black hole. This result is in agreement with idealized high-resolution binary black hole merger simulations \citep[e.g.][]{capelo2017}

So far in this section, our focus has been on the evolution of dual AGNs and their potential connection to SMBH mergers. However, dual AGNs are also suggested to be indicators of galaxy mergers \citep{comerford2009}. As mentioned earlier, we tracked dual AGNs until they resided in two distinct halos. Consistently, we found that all dual AGNs in our sample could be traced back to a previous major galaxy/halo merger. This finding aligns with earlier analyses of cosmological simulations \citep{steinborn2016,rosas-guevara2019,Volonteri2022dualhorizon,chen2023astrid} which suggest that dual AGNs are typically associated with mergers involving substantial mass ratios. Our analysis indicates that these mass ratios typically range from 10:1 to 1:1, with an average of $\sim$ 2.2:1. This fraction is slightly larger than the $\sim$ 3:1 ratio found by \citet{Volonteri2022dualhorizon} for dual AGNs in the same galaxies. The discrepancy likely arises from the different time points at which we calculate the mass ratio. We assess the mass ratio of host galaxies of dual AGNs when they are completely outside each other’s host halos $\rm R_\mathrm{200}$, whereas \citet{Volonteri2022dualhorizon} computed it when only the galaxies were separate. During a halo encounter, the smaller galaxy tends to lose mass through stripping, while the larger one gains mass which results in a smaller mass ratio.

Furthermore, linking all dual AGNs in our sample to halo/galaxy mergers, coupled with the finding that 95\% of SMBHs in these systems ultimately form binary SMBHs—with 70\% merging and 25\% remaining in binary systems—demonstrates that dual AGN systems are a more reliable indicator of binary SMBH formation and subsequent mergers compared to an average galaxy merger \citep{tremmel2018dancing}. This phenomenon is likely due to the fact that galaxy mergers resulting in dual AGNs are typically massive, major mergers.

\section{Summary and Conclusion}\label{sec:conclusion}

In this paper, we first analyzed the properties of AGNs and their neighboring SMBHs. AGNs were identified using the canonical luminosity threshold of $L_{bol} > 10^{43} $ erg s$^{-1}$while neighboring SMBHs were identified based on their distance being less than 30 kpc from an AGN, without applying any luminosity or mass thresholds. Subsequently, we applied the luminosity threshold of $L_{bol} > 10^{43} $ erg s$^{-1}$to the neighboring SMBHs, which led to the identification of dual and multiple AGNs. We then characterized the properties and evolution of these dual AGNs. All analyses are conducted at redshifts $z$ = 0.05, 0.25, 0.5, 0.75, 1, 1.22, 1.5, 1.79 and 2 using the high-resolution {\sc Romulus25} cosmological simulation. {\sc Romulus25} simulation carefully corrects the dynamical friction force onto SMBHs and produces occupation fractions consistent with observations
using only local seeding conditions. The main findings are summarized as follows:

\begin{itemize}
    \item The number of AGNs in {\sc Romulus} increases from lower to higher redshifts, which is consistent with observational studies. Consequently, the total number of neighboring SMBHs also increases.
    Every AGN has at least one neighbor SMBH. The number of these neighbors scales linearly with halo mass and can be as high as 120 neighbor SMBHs in a $10^{13} \rm M_\odot$ halo. The spatial distribution of these neighboring SMBHs also correlates linearly with their number. Thus, in higher mass halos, there are not only more neighboring SMBHs but also a more extended spatial distribution. \citet{ricarte2021wanderingbhs} find that the majority of these neighboring SMBHs originate from the centers of destroyed infalling satellite galaxies. This finding supports the expectation that higher mass halos, having undergone more mergers and also having a higher orbital decay timescale, would naturally harbor a greater number of neighboring SMBHs.

    \item Investigating the mass and luminosity range of neighboring SMBHs, We find SMBHs with luminosities exceeding $L_{bol} > 10^{43} $ erg s$^{-1}$ around primary AGNs, indicating the presence of dual and multiple AGN systems in the {\sc Romulus25} simulation.
    \vida{The occurrence of dual and multiple AGN systems, compared to single AGNs, could indicate the frequency of massive galaxy mergers and highlight potential binary SMBHs that are sources of nano-Hertz gravitational waves.}
    The number of dual AGNs increases with redshift, while multiple AGNs (triple and quadruple) are predominantly found at $z$ > 1. 
    \vida{ Our analysis of the dual AGN fraction in {\sc Romulus} shows a weakly rising trend. Although the results are noisy due to the small box size and limited statistics, they are consistent with other cosmological simulations like {\sc ASTRID}, {\sc Horizon AGN}, and {\sc EAGLE}.}

    \item 
    
    \vida{We find that the number of dual AGNs in our sample with separations of 0.5 kpc < r < 4 kpc is twice that of those with separations of 4 kpc < r < 30 kpc. This result predicts a substantial number of dual AGNs with small separations, motivating deep surveys and sensitive instruments to test this prediction.}


    \item In dual AGN systems, the SMBHs exhibit luminosity and mass ranges similar to those of single AGNs, although the differences are statistically insignificant. A notable trend is that secondary SMBHs in dual systems more frequently have masses lower than  $10^7 M_\odot$. This suggests that a low-mass black hole with a bolometric luminosity exceeding $L_{bol} > 10^{43} $ erg s$^{-1}$ is more likely to be in a dual system than to exist as a single AGN. Additionally, a larger proportion of primary SMBHs in dual systems display luminosities higher than $ 10^{44} $ erg s$^{-1}$compared to their single AGN counterparts. 

    \item The properties of host halos for dual AGNs, including halo mass, stellar mass, SFR, and gas mass, are largely consistent with those of the single AGN population, albeit marginally skewed towards the higher end. The SMBHs in dual systems tend to be undermassive relative to their host halos, exhibiting an SMBH mass to halo mass ratio below the median value found in similarly massive SMBHs. A possible explanation for this is that during halo mergers, black holes become associated with a higher mass remnant halo well before they themselves merge, thereby not keeping pace with the halo's growth mass.

   \item The analyses of dual AGN systems in the sample in this study reveal a diverse range of evolutionary patterns, including rapidly evolving systems, slower ones, and even cases where SMBH mergers are ineffective. Approximately 70\% of dual AGNs are associated with an ensuing SMBH merger, 25\% orbit in a stalled pair, and last 5\% experience ejecting from the system. The time span for mergers varies significantly, ranging from a few hundred Myr years to over 7 Gyrs. We found that duals with total bolometric luminosity $L_{bol} > 3 \times 10^{44}$ erg s$^{-1}$will merge in a timeframe of less than 4 Gyrs regardless of their separation.

    \item An investigation into the mass and luminosity ratios of SMBHs in dual AGN systems reveals instances where a lower mass SMBH exhibits a higher luminosity than its more massive counterpart. A more detailed analysis indicates that this occurs when the less massive SMBH passes through an area with higher gas density compared to the environment of the more massive SMBH, leading to an increase in its luminosity, sometimes even surpassing that of the larger black hole. This finding underscores the significance of local environmental properties in influencing the evolution of SMBHs.

    \item All dual AGN in the sample of this study are linked to a halo merger, typically characterized by a mass ratio of $\sim$ 2.2:1.

\end{itemize}

In conclusion, the findings of this paper indicate the existence of a substantial population of neighboring SMBHs and the prevalence of dual/multiple AGNs. The discovery of numerous neighboring SMBHs with luminosities slightly below the common threshold of $10^{43} $ erg s$^{-1}$calls for more sensitive instruments. With improved sensitivity, a significantly larger number of AGNs with luminous companions could be detected. Additionally, the finding of dual AGNs with sub-kpc separations underscores the need for both higher sensitivity instruments and higher resolution cosmological simulations. This dual approach is essential for the deeper study of these systems, both observationally and theoretically.

\section*{Acknowledgements}

\vida{We thank the anonymous referee for their very useful comments.} We also thank Pedro R. Capelo for insightful discussions and suggestions. 
VS and AB acknowledge support from the Natural Sciences and Engineering Research Council of Canada (NSERC) through its Discovery Grant program. AB acknowledges support from the Infosys Foundation via an endowed Infosys Visiting Chair Professorship at the Indian Institute of Science. \vida {AB also acknowledges support from the Leverhulme Trust via the Leverhulme visiting professorship at the University of Edinburgh} The work of SM is a part of the $\langle \texttt{data|theory}\rangle$ \texttt{Universe-Lab} which is supported by the TIFR and the Department of Atomic Energy, Government of India. MT was supported by an NSF Astronomy and Astrophysics Postdoctoral Fellowship under award AST-2001810. AB, TQ, and MT were partially supported by NSF award AST-1514868. 

The {\sc Romulus} simulation suite is part of the Blue Waters sustained-petascale computing project, which is supported by the National Science Foundation (via awards OCI-0725070, ACI-1238993, and OAC-1613674) and the state of Illinois. Blue Waters is a joint effort of the University of Illinois at Urbana-Champaign and its National Center for Supercomputing Applications. Resources supporting this work were also provided by the (a) NASA High-End Computing (HEC) Program through the NASA Advanced Supercomputing (NAS) Division at Ames Research Center; and (b) Extreme Science and Engineering Discovery Environment (XSEDE), supported by National Science Foundation grant number ACI-1548562. The analysis reported in this paper was enabled in part by WestGrid and Digital Research Alliance of Canada (alliancecan.ca) and on the cluster of $\langle \texttt{data|theory}\rangle$ \texttt{Universe-Lab} supported by DAE. Our analysis was performed using the Python programming language (Python Software Foundation, https://www.python.org). The following packages were used throughout the analysis: numpy (\citealt{harris2020array}), matplotlib (\citealt{hunter2007matplotlib}), Pynbody (\citealt{pontzen2013pynbody}), SciPy (\citealt{virtanen2020scipy}), and TANGOS (\citealt{pontzen2018tangos}).

Finally, VS and AB acknowledge the l{\fontencoding{T4}\selectfont
\M{e}}\'{k}$^{\rm w}${\fontencoding{T4}\selectfont\M{e}\m{n}\M{e}}n 
peoples on whose traditional territory the University of Victoria stands, and the Songhees, Equimalt and
\b{W}S\'{A}NE\'{C} peoples whose historical relationships with the land continue to this day.



\section*{Data Availability}

The data directly related to this article will be shared on reasonable request to the corresponding author. Galaxy database and particle data for {\sc Romulus} is available upon request from Michael Tremmel.

\bibliography{All}{}
\bibliographystyle{aasjournal}



\end{document}